\documentclass{emulateapj}

\begin{document}

\title{GOODS-{\em Herschel}: Separating High Redshift active galactic Nuclei and star forming galaxies Using Infrared Color Diagnostics}
\slugcomment{to be submitted to ApJ}

\author{Allison Kirkpatrick\altaffilmark{1}, Alexandra Pope\altaffilmark{1}, Vassilis Charmandaris\altaffilmark{2,3,4}, Emmanuele Daddi\altaffilmark{5},  David Elbaz\altaffilmark{6}, Ho Seong Hwang\altaffilmark{7}, 
Maurilio Pannella\altaffilmark{6}, Douglas Scott\altaffilmark{8},
Bruno Altieri\altaffilmark{9}, Herve Aussel\altaffilmark{6}, Daniela Coia\altaffilmark{9}, Helmut Dannerbauer\altaffilmark{10}, Kalliopi Dasyra\altaffilmark{6}, Mark Dickinson\altaffilmark{11},
Jeyhan Kartaltepe\altaffilmark{11}, Roger Leiton\altaffilmark{6,12}, Georgios Magdis\altaffilmark{13}, Benjamin Magnelli\altaffilmark{14}, Paola Popesso\altaffilmark{14},
Ivan Valtchanov\altaffilmark{9}}
\altaffiltext{1}{Department of Astronomy, University of Massachusetts, Amherst, MA 01002}
\altaffiltext{2}{Department of Physics and Institute of Theoretical \& Computational Physics, University of Crete, GR-71003, Heraklion, Greece}
\altaffiltext{3}{IESL/Foundation for Research \& Technology-Hellas, GR-71110, Heraklion, Greece}
\altaffiltext{4}{Chercheur Associ\'e, Observatoire de Paris, F-75014,  Paris, France}
\altaffiltext{5}{Laboratoire AIM, CEA/DSM-CNRS-Universit{\'e} Paris Diderot, Irfu/SAp, Orme des Merisiers, F-91191 Gif-sur-Yvette, France}
\altaffiltext{6}{Laboratoire AIM, CEA/DSM-CNRS-Universit{\'e} Paris Diderot, Irfu/SAp, Orme des Merisiers, F-91191 Gif-sur-Yvette, France}
\altaffiltext{7}{Smithsonian Astrophysical Observatory, 60 Garden Street, Cambridge, MA, 02138, USA}
\altaffiltext{8}{Max-Planck-Institut f\"ur Extraterrestrische Physik (MPE), Postfach 1312, 85741, Garching, Germany}
\altaffiltext{9}{Herschel Science Centre, European Space Astronomy Centre, Villanueva de la Ca\~nada, 28691 Madrid, Spain}
\altaffiltext{10}{Universit\"at Wien, Institut f\"ur Astrophysik, T\"urkenschanzstra\ss e 17, 1180 Wien, Austria}
\altaffiltext{11}{National Optical Astronomy Observatory, 950 North Cherry Avenue, Tucson, AZ 85719, USA}
\altaffiltext{12}{Astronomy Department, Universidad de Concepci\'on, Casilla 160-C, Concepci\'on, Chile}
\altaffiltext{13}{Department of Physics, University of Oxford, Keble Road, Oxford OX1 3RH, UK}
\altaffiltext{14}{Max-Planck-Institut f\"ur Extraterrestrische Physik (MPE), Postfach 1312, 85741, Garching, Germany}


\begin{abstract}
We have compiled a large sample of 151 high redshift ($z=0.5-4$) galaxies selected at 24$\,\mu$m ($S_{24}>100\,\mu$Jy) in the GOODS-N and ECDFS fields for which we have deep {\em Spitzer} IRS spectroscopy,
allowing us to decompose the mid-infrared spectrum into contributions from star formation and activity in the galactic nuclei.
In addition, we have a wealth of photometric data from {\em Spitzer} IRAC/MIPS and
{\em Herschel} PACS/SPIRE. We explore how effective different infrared color combinations are at separating our mid-IR spectroscopically determined active galactic nuclei from our star forming galaxies. We look in depth
at existing IRAC color diagnostics, and we explore new color-color diagnostics combining mid-IR, far-IR, and near-IR photometry,
since these combinations provide the most detail about the shape of a source's IR spectrum. An added benefit of using a color that combines far-IR and mid-IR photometry is that it is indicative of the power source driving the IR luminosity. For our data set, the optimal color selections are $S_{250}/S_{24}$ vs. $S_{8}/S_{3.6}$ and $S_{100}/S_{24}$ vs. $S_{8}/S_{3.6}$; both diagnostics have $\sim$10\% contamination rate in the
regions occupied primarily by star forming galaxies and active galactic nuclei, respectively.
Based on the low contamination rate, these two new IR color-color diagnostics are ideal for estimating both the mid-IR power source of a galaxy when spectroscopy is unavailable and the dominant power source contributing to the IR luminosity.
In the absence of far-IR data, we present color diagnostics using the WISE mid-IR bands which can efficiently select out high $z$ ($z\sim2$) star forming galaxies.
\end{abstract}

\section{Introduction}
In the current narrative of galaxy evolution, star formation and the growth of supermassive black holes are intertwined.
The star formation rate density of the Universe peaks from $z\sim1-3$ \citep[e.g.,][]{bouwens2009,magnelli2011,murphy2011}, an epoch in which the black holes within the center of massive galaxies are 
simultaneously building up their mass \citep{wall2005,kelly2010}. There is compelling evidence that the growth of black holes and the buildup of stellar mass is linked, though the processes
that regulate, and ultimately quench, the simultaneous growth of the stellar mass and black hole are not yet fully disentagled \citep[e.g.,][]{mullaney2012}. 

To study the properties of active galactic nuclei (AGN) and star-forming galaxies at $z\sim1-3$, when the star formation rate in the most massive galaxies begins to decline,
it is necessary to first identify systems likely harboring an AGN.
Because AGN are much more luminous in the X-ray than star-forming galaxies, X-ray detection provides one of the best means of identifying an AGN \citep[e.g.,][]{alexander2003}.
Deep field surveys with the {\it Chandra X-ray Observatory} have exposed an AGN population out to $z\sim5$
\citep[e.g.,][]{brandt2001,giacconi2002}. However, detailed X-ray spectral analysis shows that the majority of sources are obscured by gas and dust (see \citealp{brandt2005} for a review), and
there is likely a sizeable fraction of AGN that remain undetected by {\it Chandra} surveys, as evidenced by the fact that nearly half of the X-ray background
is unresolved at $>$\,6 keV \citep{worsley2005}, and the ratio of observed obscured to unobscured AGN at high redshifts is lower than what is found for comparably luminous AGN in the local Universe
\citep[e.g.,][]{treister2005}.

Due to the incompleteness of surveys conducted at X-ray wavelengths, we must employ alternate methods insensitive to dust obscuration to identify the presence of an AGN. The infrared portion of the 
spectral energy distribution (SED) shows clear signs of both AGN and star formation (SF) activity. Based on the presence of polycyclic aromatic hydrocarbons and continuum thermal dust emission,
mid-infrared spectra can be decomposed into the relative contributions of SF and AGN activity \citep[e.g.,][]{laurent2000,armus2007,sajina2007,pope2008,kirkpatrick2012}. At $1.6\,\mu$m, SF dominated galaxies
will exhibit a stellar bump due to emission from older stellar populations, whereas as the dust surrounding luminous AGN heats up, it will radiate into the near-IR and mid-IR, creating a pure
power-law spectrum at these wavelengths. Finally, the peak and shape of the far-IR emission depends on the temperature of the dust, which becomes warmer as the AGN grows more luminous \citep[e.g.,][]{haas2003,sanders1988}. Due to
AGN signatures from the near-IR to the far-IR, infrared color-selection techniques are a promising way to select out AGN missed by X-ray surveys, or when X-ray data are unavailable.

In this work, we explore the IR color space for a sample of 151 high redshift luminous infrared galaxies (LIRGs, $L_{\rm IR} = 10^{11}-10^{12}\,{\rm L}_{\odot}$) and ultra luminous infrared galaxies (ULIRGs, $L_{\rm IR} > 10^{12}\,{\rm L}_{\odot}$) at high redshift ($z\sim0.5-3.5$) with deep {\em Spitzer} mid-IR spectroscopy and a suite of multiwavelength photometry spanning $3.6 - 500\,\mu$m. Several studies have explored using {\it Spitzer} IRAC colors as a means to separate AGN \citep{lacy2004,stern2005,donley2012}. For the first time, we explore these various diagnostics using a sample of mid-IR spectroscopically determined AGN and SF galaxies. In light of the emerging Wide-field Infrared Survey Explorer (WISE) photometry, 
we also discuss how well mid-IR diagnostics separate our spectroscopic AGN and SF sources. The focus of the paper, however, lies in combining near-, mid-, and far-IR photometry
to distinguish our AGN from the SF galaxies. The far-IR is tracing the bulk of the IR luminosity, and if a galaxy's mid-IR power source is also affecting the far-IR, we expect that
combining both portions of the spectrum will be a useful diagnostic.

We build on a previous paper \citep[hereafter Paper I]{kirkpatrick2012} in which we separate
our sample into AGN and SF dominated, based on mid-IR spectral decomposition, and analyze the average IR SEDs of each, including mid-IR spectral features, IR luminosity, and dust temperatures. We have spectroscopically
determined the nature of the mid-IR power source for our individual sources, and we have created SEDs that represent the average features of our SF and AGN sources. Now we look for advantageous color-color cuts that can be used to select out AGN candidates when spectroscopy is unavailable. We explore different combinations of colors
using both the photometry of our individual sources, as well as redshift tracks from the composite SEDs, to determine color combinations that separate AGN from SF galaxies.
 
The paper is laid out as follows: in Section~2, we describe details of our sample, the mid-IR spectral decomposition of individual sources, and the composite spectral energy distributions
(SEDs) we create from this sample. In Section~3, we explore the efficacy of existing IRAC color selection techniques for high redshift sources. In Section~4, we present new color diagnostics based on
combining photometry from the {\em Spitzer Space Telescope} and the {\em Herschel Space Observatory}. We apply our new diagnostics to the broader GOODS-N and ECDFS fields in Section~5, and we end with our conclusions in Section~6.
Throughout this paper, we assume a standard cosmology with $H_{0}=70\,\rm{km}\,\rm{s}^{-1}\,\rm{Mpc}^{-1}$, $\Omega_{\rm{M}}=0.3$ and $\Omega_{\Lambda}=0.7$. 

\section{Data}
A full description of our sample and the composite SEDs we create from it is given in Paper I. Here, we summarize the main details and results.

\subsection{Multiwavelength data}
Our sample consists of 151 high redshift galaxies from the Great Observatories Origins Deep Survey North (GOODS-N) and Extended Chandra Deep Field Survey (ECDFS) fields. We include all sources in these fields that were observed with the {\it Spitzer} IRS. 
While this sample contains a diverse range of sources depending on the goals of each individual observing program, the overlying selection criteria is that each source must be detected at 24$\,\mu$m with a flux of $S_{24}\gtrsim100\,\mu$Jy, since anything fainter will not be observable with the IRS in less than 10 hours. More details on this database of IRS sources in GOODS-N and ECDFS can be found in our data paper (Pope et al.~in preparation). 

The GOODS fields have been extensively surveyed and are rich in
deep multiwavelength data including: ground-based imaging in the near-IR ($J$ and $K$ bands) from VLT/ISAAC \citep{retzlaff2010} and CFHT/WIRCAM \citep{wang2010,lin2012}; {\em Chandra} 2 Ms X-ray observations \citep{alexander2003, luo2008}; 3.6, 4.5, 5.8, 8.0\,$\mu$m from the Infrared Array Camera (IRAC) on {\em Sptizer};
IRS peak-up observations at $16\,\mu$m \citep{teplitz2011} and MIPS imaging at 24 and $70\,\mu$m \citep{magnelli2011}. Recently, GOODS-N and GOODS-S have been surveyed with the GOODS-{\em Herschel} Open Time Key Program
\citep[P.I. David Elbaz,][]{elbaz2011} using both the PACS and SPIRE
instruments providing deep photometry at five far-IR wavelengths: 100, 160, 250, 350, and $500\,\mu$m. For the present study, we combine space-based imaging from {\em Spitzer} and {\em Herschel}
to obtain 12 photometric bandwidths spanning the near-IR to the far-IR.

For sources lacking a detection at the {\it Herschel} wavelengths, we extract a measurement of the flux density and associated uncertainty for each source directly from our images.
The images are in units of mJy/beam, so we use the 24 $\mu$m prior positions to find the appropriate pixels for each galaxy. We do not take a measurement when a source looks too blended on the image itself. In Paper I, we present complete {\em Herschel} photometry for our IRS sources, as well as indicating which sources have measurements and
which have detections at the {\em Herschel} wavelengths.

We performed spectral decomposition of the {\em Spitzer} IRS mid-IR spectrum for each source in order to disentangle the AGN and SF components.
We follow the technique outlined in detail in \citet{pope2008} which we summarize here. We fit the individual spectra with a
model comprised of three components: (1) The SF component is represented by either the local starburst composite of \citet{brandl2006} or simply the mid-IR spectrum of the prototypical 
starburst M82 \citep{schreiber2003} -- with the SNR, wavelength coverage and spectral resolution of our high redshift spectra both give equally good fits to the SF component of our galaxies.
(2) The AGN component is determined by fitting a pure power-law with the slope and normalization
as free parameters. (3) An extinction curve from the \citet{draine2003} dust models is applied to the AGN component. The extinction curve is not monotonic in wavelength
and contains silicate absorption features, the most notable for our wavelength range being at $9.7\,\mu$m. We fit all three components simultaneously and integrate under the PAH and continuum components to determine the fraction of the mid-IR luminosity ($\sim5-12\,\mu$m) from SF and AGN activity, respectively.
For each source, we quantify the strength of the AGN in terms of the percentage of the total mid-IR luminosity coming from the AGN continuum component. Based on this mid-IR spectral decomposition, we find that 38 (25\%) out of our sample of 151 galaxies are dominated ($\ge$~50\% of luminosity) in the mid-IR by an AGN.

\begin{figure}
\plotone{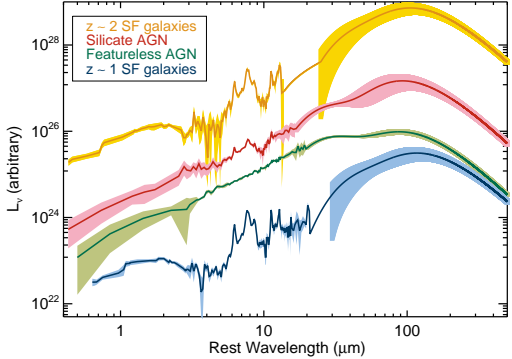}
\caption{Composite SEDs (and associated uncertainties) for each of our subsamples (see Table \ref{basictbl}) created by stacking photometry and spectroscopy in the near- and mid-IR and fitting a two-temperature model to the far-IR. The composites have been offset to allow for easy comparison. The SF templates lack uncertainties in the range $\sim18-30\,\mu$m as we lacked data in this wavelength range, and so we interpolate between the mid-IR and far-IR portions of the spectra. Composite SEDs are publicly available. \label{composites}}
\end{figure}

\subsection{Galaxy Classifications}
To more thoroughly compare the mid-IR spectral properties and full IR SEDs within our sample, we divide our sample into four primary subsamples based on the results of the mid-IR spectral decomposition. First, each galaxy is classified as
either SF- or AGN- dominated based on having $<50\%$ and $>50\%$ AGN contribution in the mid-IR, respectively. We further divide the SF galaxies into two bins: $z\sim1$ ($z<1.5$) and $z\sim2$ ($z>1.5$). The AGN sources are likewise separated into two bins: those with measurable $9.7\,\mu$m silicate absorption (hereafter referred to as silicate AGN) and those without (hereafter referred to as featureless AGN). We are unable to further classify four AGN sources as they lack spectral coverage in the relevant range ($9-10\,\mu$m) to determine whether they exhibit silicate absorption. We refer to these as unclassifiable AGN in the relevant figures. Our four primary subsamples are listed in Table \ref{basictbl} along with their median redshifts and observed frame median 24, 100, and 8$\,\mu$m flux densities. 

\begin{deluxetable*}{lrlrlcrlrlrl}[h!]
\renewcommand{\tabcolsep}{0pt}
\tablecolumns{12}
\tablecaption{Basic properties of our four sub-samples. \label{basictbl}}


\tablehead{\colhead{} & \multicolumn{2}{c}{\# of\,\tablenotemark{a}} & \multicolumn{2}{c}{Median (Mean)} & \colhead{$L_{\rm IR}$\tablenotemark{b}} & 
\multicolumn{2}{c}{Median $S_{24}$\tablenotemark{c}} & \multicolumn{2}{c}{Median $S_{100}$ \tablenotemark{c}} &\multicolumn{2}{c}{Median $S_8$  \tablenotemark{c}} \\
\colhead{Subsample} & \multicolumn{2}{c}{Sources} & \multicolumn{2}{c}{Redshift} & \colhead{($10^{12}\,{\rm L}_\odot$)} & \multicolumn{2}{c}{($\mu$Jy)} & \multicolumn{2}{c}{(mJy)} & \multicolumn{2}{c}{($\mu$Jy)}}
\startdata
$z\sim1$ SF galaxies & 69 & (39) & 1.0 (0.9) & [0.8, 1.0] & 0.42 $\pm$ 0.17 & \ 370 & [260, 570]  &  7.9 & [4.8, 14.7]  & \ 42 & [23, 57] \\
$z\sim2$ SF galaxies & 44 & (30) & 1.9 (1.9) & [1.8, 2.1] & 2.00 $\pm$ 0.71 & \ 270 & [220, 370] &   3.2  &[1.6,\ 5.0]    & \ 17 & [13, 38] \\
Silicate AGN 	   & 22 & (17) & 1.9 (1.7) & [1.6, 2.0] & 1.65 $\pm$ 0.54 & \ 470 & [250, 860] &   5.3 & [3.0, 10.2]  & \ 64 & [21, 140] \\
Featureless AGN    & 12 & (9) & 1.2 (1.1) & [0.6, 1.6] & 0.76 $\pm$ 0.07 &  1520 & [1240, 2300]  &   9.5 & [4.0, 11.7]  & 288 & [240, 311]
\enddata
\tablenotetext{}{We list the upper and lower quartile values in brackets next to each calculated median.}
\tablenotetext{a}{We list the number of sources in each sub-sample that are used to create the composite SEDs in parentheses. We do not include sources with an incomplete IR SED
when creating the composites to avoid biasing our results (See \S \ref{sec:sed}).}
\tablenotetext{b}{Calculate by integrating under the composite SEDs from $8-1000\,\mu$m. See Paper I for details.}
\tablenotetext{c}{Observed frame fluxes.}
\end{deluxetable*}

While the majority of our sources that are classified as SF-dominated based on the mid-IR spectra have a negligible ($<\,20\%$) contribution from an AGN, the AGN dominated sources exhibit varying degrees of concurrent SF activity.
The featureless AGN (lacking silicate absorption) primarily have a very strong AGN continuum accounting for $80-100\%$ of the mid-IR emission, whereas the silicate AGN have a more uniform distribution of AGN fraction ($50-100\%$) with some silicate AGN also having weak PAH features.

\begin{figure*}[ht!]
\includegraphics[scale=0.4]{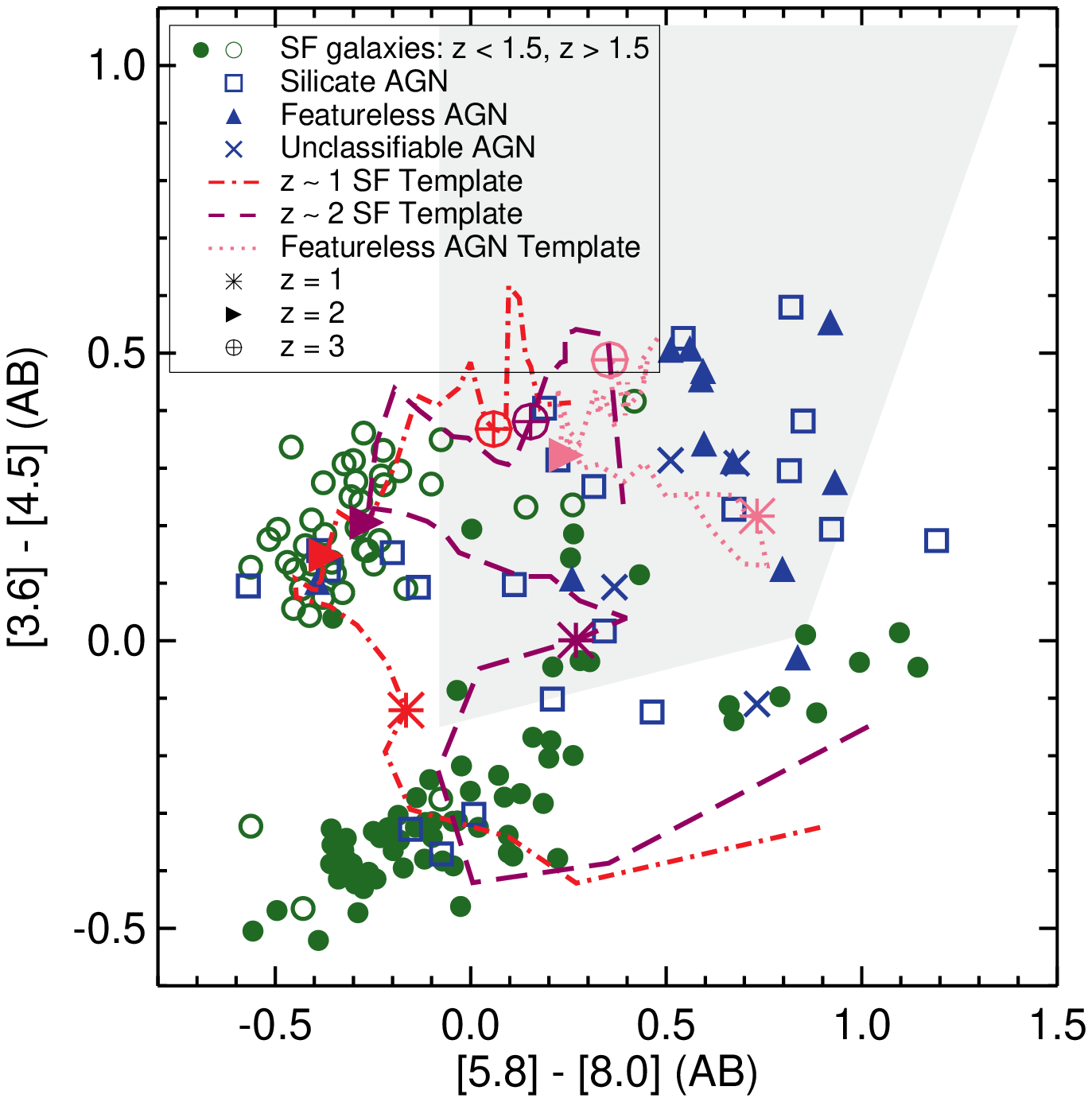}
\includegraphics[scale=0.4]{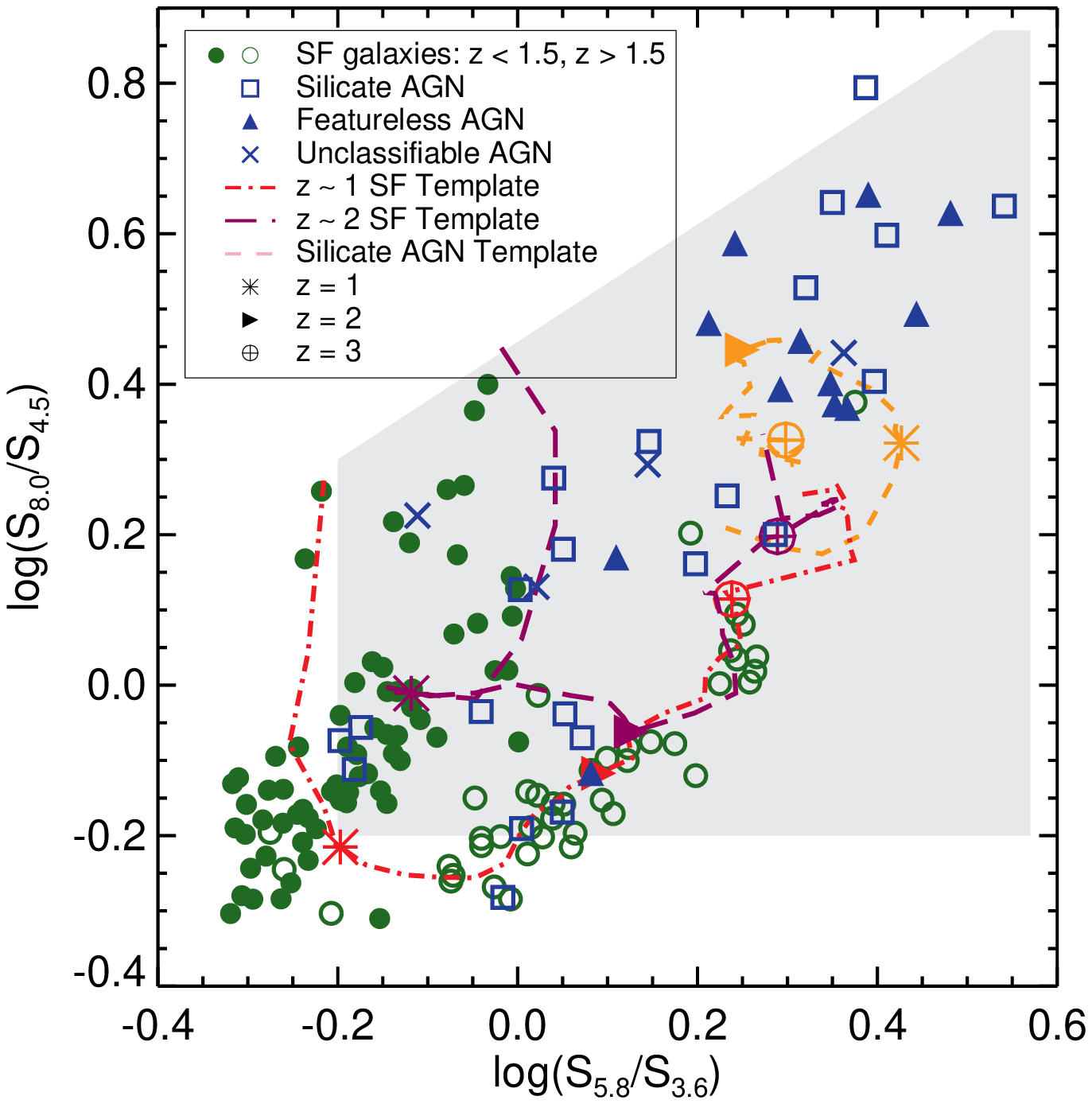}
\includegraphics[scale=0.4]{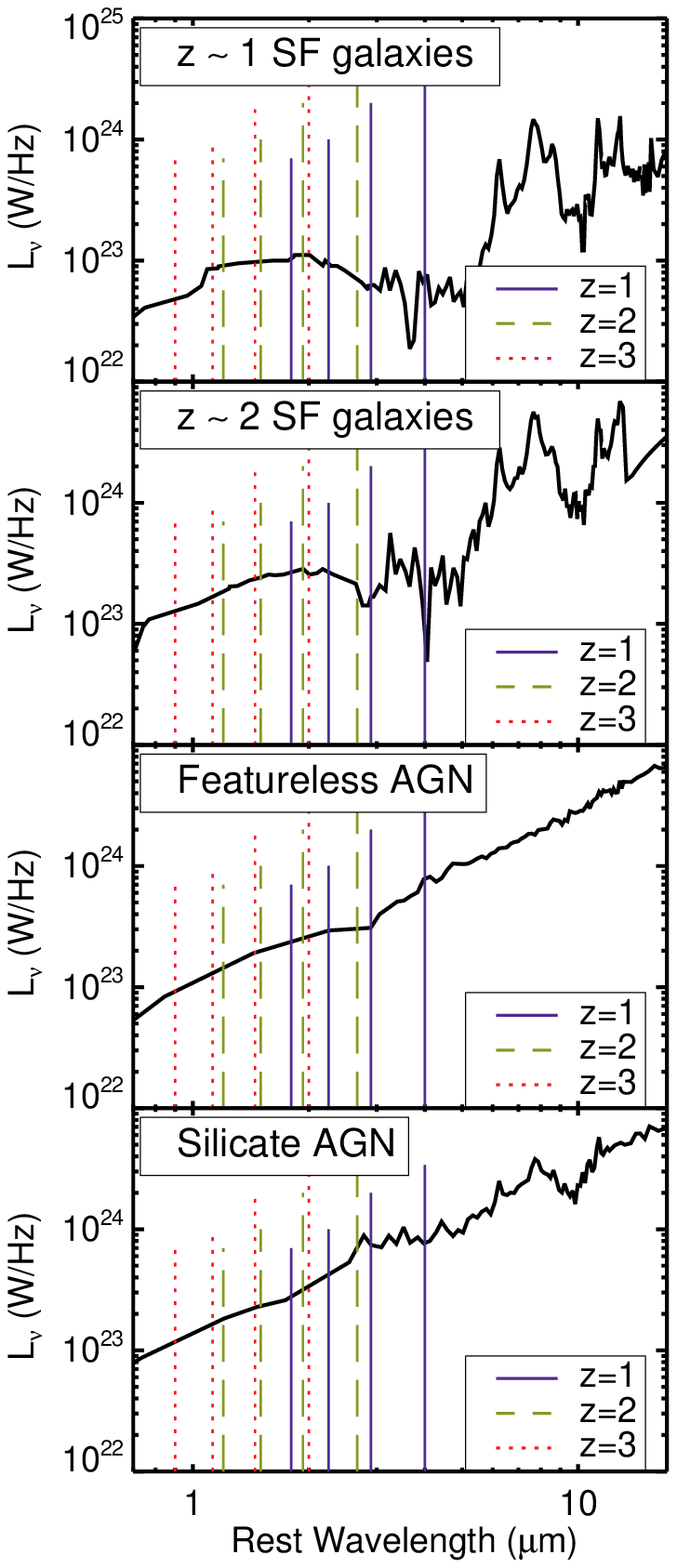}
\caption{Color-color diagnostics from \citet[{\em, left panel}]{stern2005} and \citet[{\em, middle panel}]{lacy2004} for selecting AGN dominated sources (shaded regions). 
Our SF galaxies separate nicely with redshift ($z < 1.5$ are the filled green circles while $z > 1.5$ are plotted as
the open green circles; see the online version for color figures). The AGN sources do not separate in these color-color spaces with redshift and thus are only plotted according to whether they possess a $9.7\,\mu$m absorption feature (open blue squares) or not (filled blue triangles). The four AGN we were unable to classify as they lacked spectral coverage at 9.7$\,\mu$m are
plotted as the blue crosses. Overplotted are the redshift tracks calculated from our composite SEDs for $z\sim2$ SF galaxies (purple) and $z\sim1$ SF galaxies (red). In the left panel, the $z\sim1$ SF SED and
$z\sim2$ SF SED diverge around $z\sim1$ due to the difference in the strength of portion of the stellar bump traced by the 3.6\,$\mu$m filter.
As the two AGN composites occupy the same region of colorspace for both the left and middle panels, we only plot the redshift track of the featureless AGN composite SED (in pink) on the left and the track
for the silicate absorption AGN (orange) in the middle. The right panel shows the effective wavelengths of the four IRAC bands at different redshifts on each of our composite SEDs, demonstrating that at high redshift these colors 
sample the stellar bump region. \label{sternlacy}}
\end{figure*}

\subsection{Composite SEDs}
\label{sec:sed}
To assess the average properties of our primary four subsamples of galaxies (due to the small number, we do not examine the unclassifiable AGN in detail),
we created composite SEDs from $0.3-600\,\mu$m rest-frame
by combining data from ground-based near-IR; {\em Spitzer} IRAC, IRS, and MIPS (24\,\micron, 70\,\micron); {\em Herschel} PACS and
SPIRE. 
We reject any source with less than 3 measurements or detections in the far-IR bandwidths, or any source without mid-IR
data in the range $6.4-7.5\,\mu$m, as this is the range used to normalize the individual SEDs.
After normalization, we stacked flux densities in bin sizes of $\sim0.1\,\mu$m below $20\, \mu$m. Above $20\,\mu$m, we fit all data points with a two-temperature modified blackbody.
Full details of these composites are found in Paper I.
The composites are shown in Figure \ref{composites} with each composite offset on the y-axis to allow for easier comparisons. Composite SEDs are publicly available\footnote{http://www.astro.umass.edu/$\sim$pope/Kirkpatrick2012/}.

The two SF composites are remarkably similar in shape and have most emission from cold dust. The featureless AGN composite is nearly a pure power-law until $\sim20\,\micron$, and then is relatively flat from $\sim 20 - 100\, \micron$.
The silicate AGN SED is a power-law in the near-IR, has weak PAH features and silicate absorption at $9.7\,\mu$m, has warm dust emission around $20\,\micron$, and has a cold dust component peaking at the same
wavelengths as the SF SEDs. The difference in shapes between the two AGN SEDs and the SF SEDs suggests that IR color diagnostics could be useful for separating AGN from SF galaxies.

\section{IRAC Color-Color Diagnostics}
\label{sec:irac}
Some of the most well used IR color diagnostics for separating AGN and SF galaxies are presented in \citet{lacy2004} and \citet{stern2005} and utilize IRAC colors. The motivation behind an IRAC selection technique is that, at
these wavelengths, luminous AGN should have a monotonically increasing SED, and these power-law colors will separate AGN from SF galaxies
in colorspace.
With our high redshift photometry and spectroscopy, we are able to apply these diagnostics to a large sample of mid-IR spectroscopically determined AGN and SF galaxies.
 \citet{lacy2004} and \citet{stern2005} define IRAC color-color regions to separate AGN based on large surveys of low redshift ($z \lesssim 0.7$) galaxies.
As we move to higher redshift ($z\sim2$), the IRAC bandwidths begin to probe the stellar bump;
our composite AGN and SF SEDs exhibit different shapes in the $\lambda < 4\,\micron$ region of the spectrum, 
indicating that IRAC color diagnostics might also be useful at higher redshift.
We apply the diagnostics of \citet{stern2005} and \citet{lacy2004} to our sample (Figure \ref{sternlacy}). Sources in our sample that we determined through mid-IR spectral decomposition to be dominated
by an AGN ($>50\%$) in the mid-IR are plotted in blue (see the online version for color figures), and sources dominated by star formation are plotted in green.

\begin{figure}
\plotone{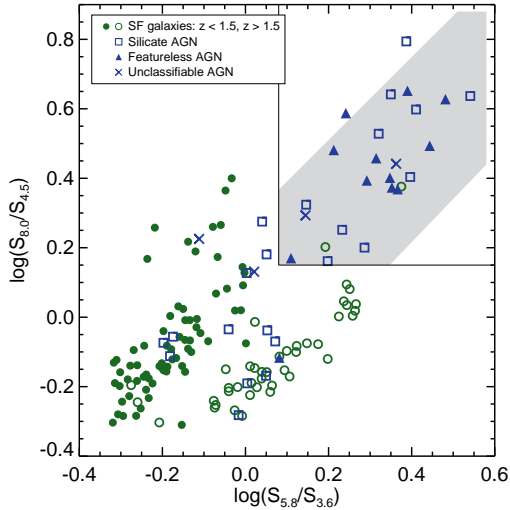}
\caption{Our sources in IRAC color space with the criteria of \citet{donley2012} shown as the shaded region. The modified criteria do an effective job of selecting power-law sources while avoiding contamination from SF galaxies but still miss many of our silicate AGN.\label{donley}}
\end{figure}

In the left panel of Fig. \ref{sternlacy}, we plot our sources in the IRAC color-color space defined in \citet{stern2005} with the gray shaded region being the area defined
as AGN-dominated by the authors.
Our SF sources separate cleanly according to redshift and largely avoid the gray shaded region. We have also overplotted the redshift tracks of our $z\sim1$ SF composite SED and our $z\sim2$ SF composite SED. We calculate the 
redshift tracks of each of our composite SEDs by convolving the SEDs with the IRAC bandpass filters (and MIPS, PACS, SPIRE, and WISE filters in \S \ref{sec:new}) at the appropriate wavelengths for a given 
redshift. The convolution acts to smooth out much of the noise in our spectra as the individual filters span a fairly large wavelength range.
The $z\sim2$ SF track contaminates the AGN region around $z\sim1$, and in fact, all of the $z\sim1$ SF galaxies
(green filled circles) occupying the shaded region have a redshift between 1 and 1.5. The divergence of the two SF tracks around $z\sim1$ is due to the fact that the $3.6\,\mu$m filter is tracing the bluest portion of 
the stellar bump, which has differing strengths relative to the other IRAC filters for the two SF composite spectra (see Fig. \ref{composites} and the right panel of Fig. \ref{sternlacy} ). Our individual
$z\sim1$ sources have colors that are located around both the $z\sim1$ and $z\sim2$ tracks indicating a spread in the obscuration of the IRAC colors for these sources. The spread of the individual photometry points in relation to the composite SED is illustrated in Paper I.

\begin{figure*}[ht!]
\includegraphics[scale=0.37]{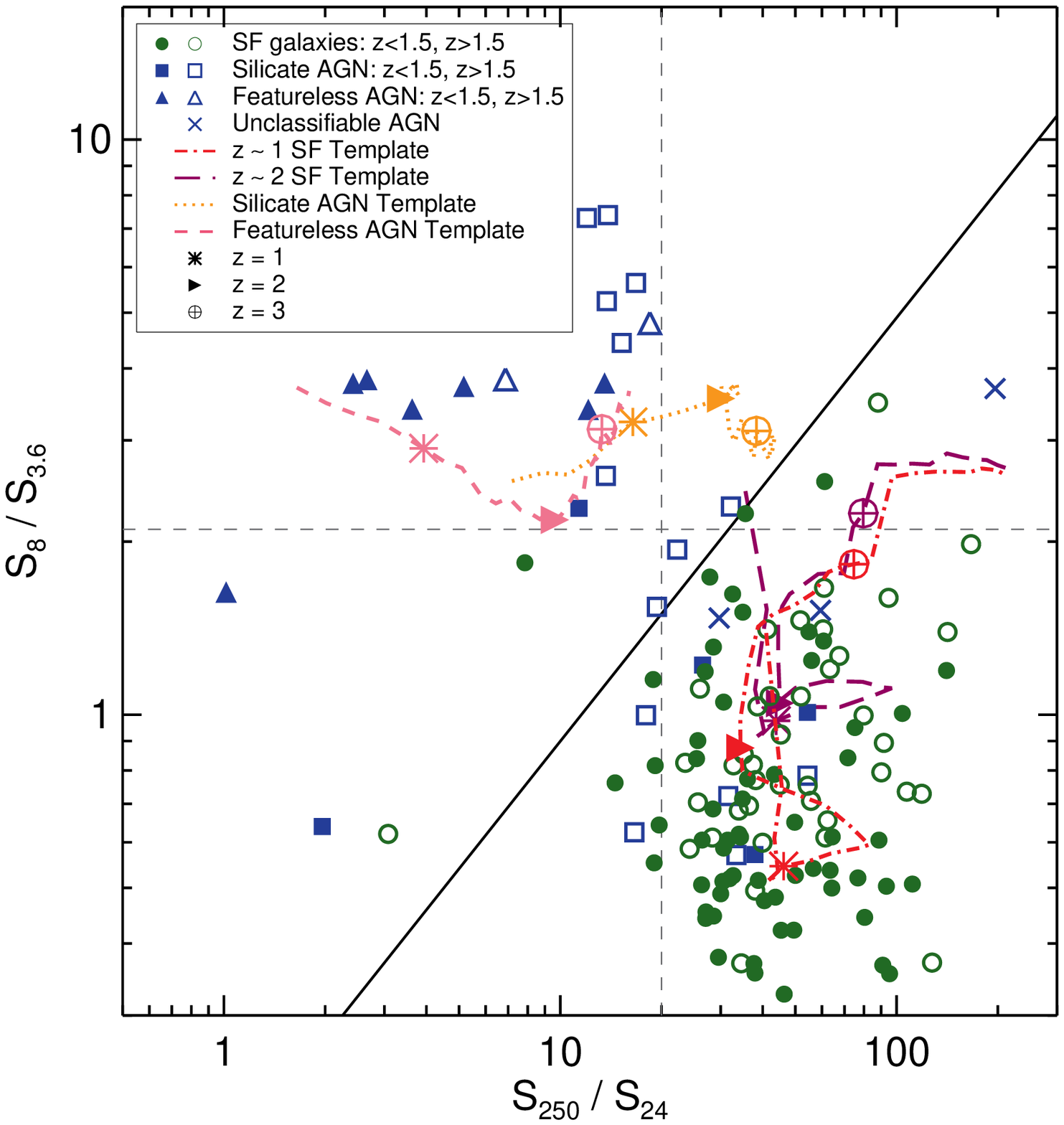}
\includegraphics[scale=0.365]{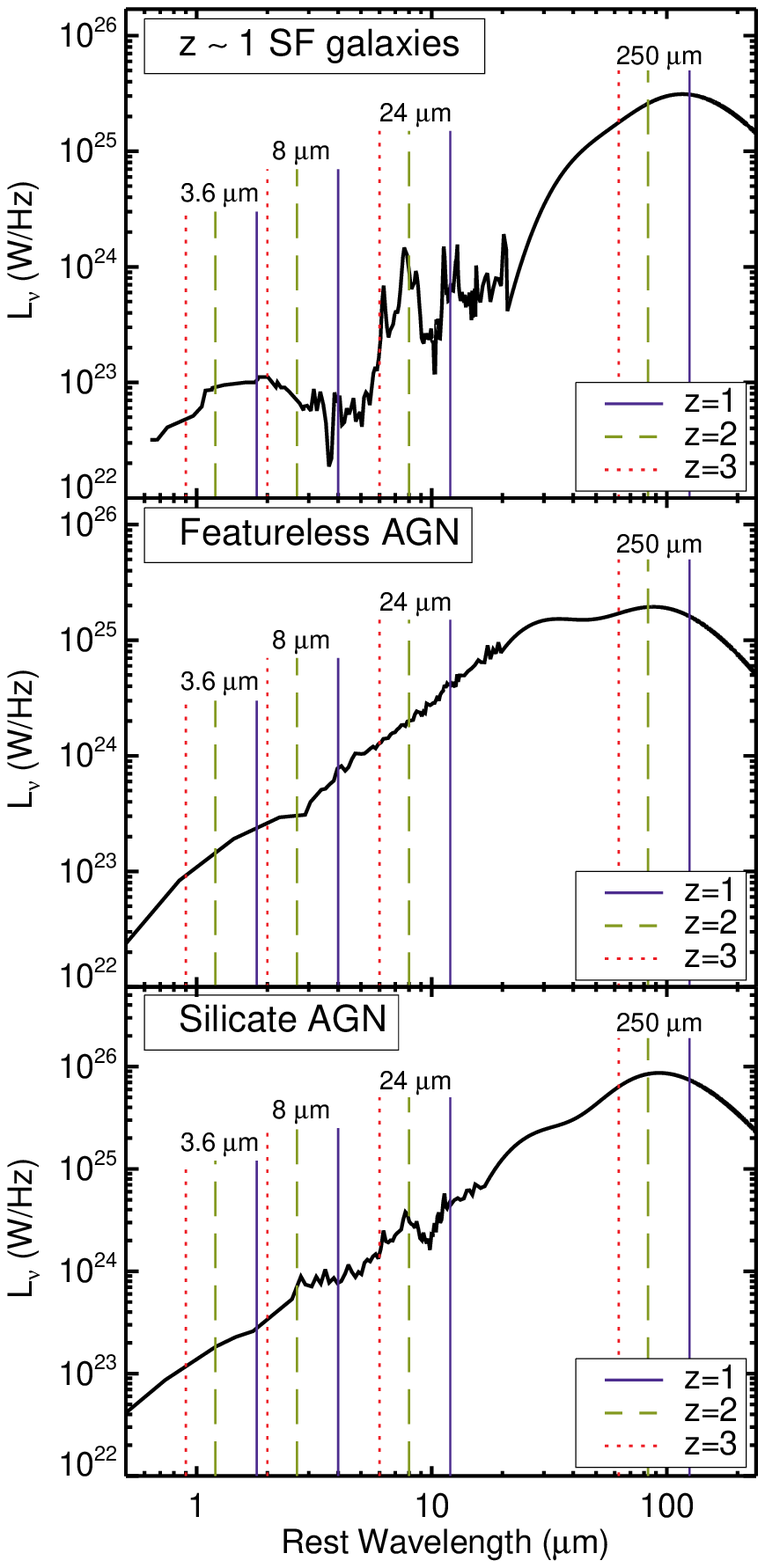}
\includegraphics[scale=0.37]{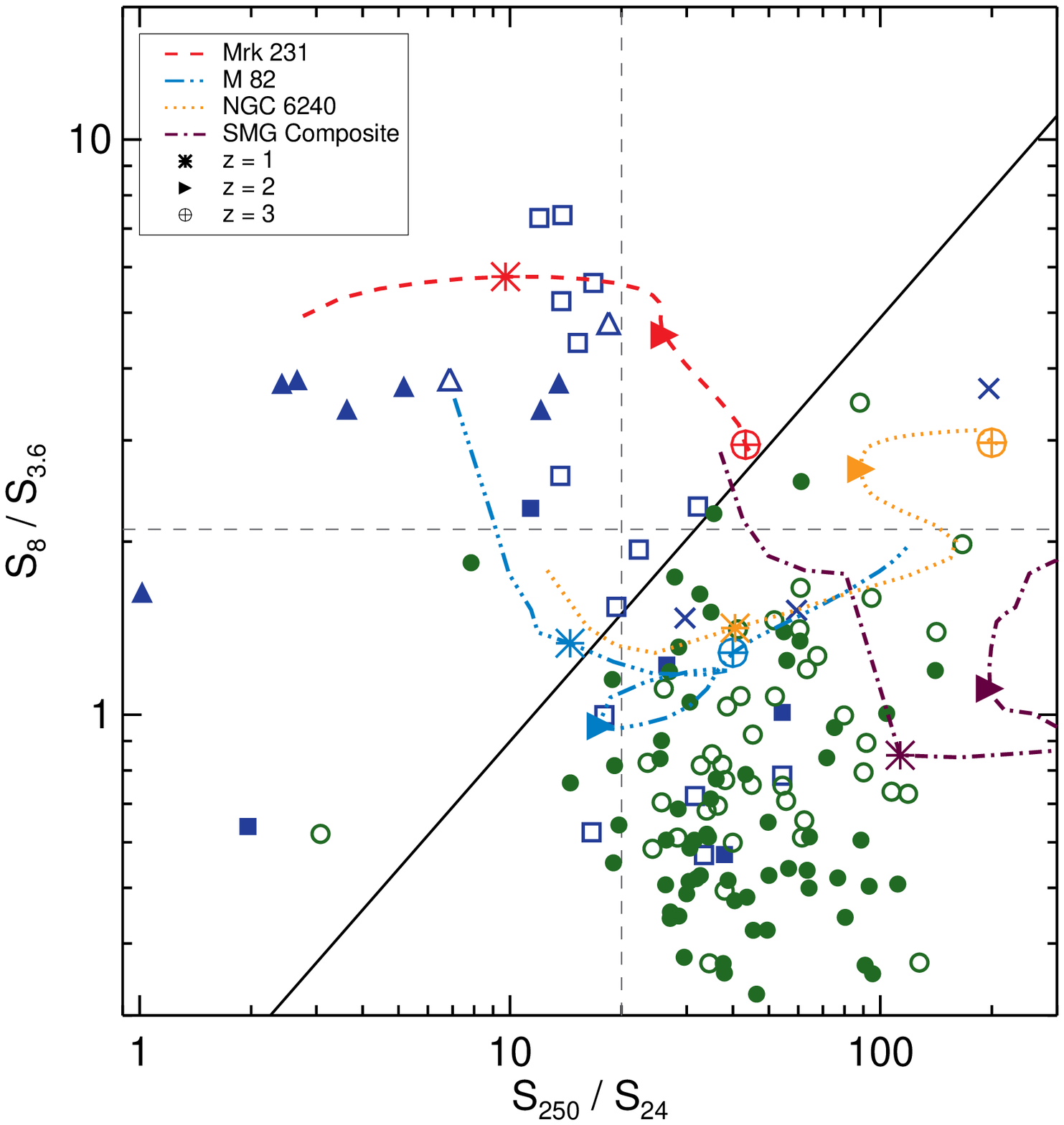}
\caption{{\em Left -} New color-color diagnostic combining submm wavelength {\it Herschel}/SPIRE photometry with mid-IR {\em Spitzer} data. The dark line divides the SF galaxies (green) from the AGN galaxies (blue).
We plot all sources according to redshift (filled symbol: $z<1.5$; open symbol: $z>1.5$).  We overplot the redshift tracks of our composite SEDs from $z\sim0.5-4.0$ as this is the redshift range of our sources.
We show with the dashed grey lines what color thresholds can be used
to separate AGN from SF sources if only one color is available.
{\em Middle -} Illustrates where each of the redshifted photometry points lie on the composite SEDs. 
{\em Right -} Same as the left panel except we overplot the SED templates of local galaxies and high redshift SMGs for comparison. 
\label{basic250_24}}
\end{figure*}

The AGN sources (blue symbols) are not completely constrained to the gray shaded region, though our redshift track from the featureless AGN composite SED indicates they should be on average, illustrating a lack of homogeneity in the near- and mid-IR photometry among mid-IR classified AGN.
Most AGN in our sample do not have a clear stellar bump in the near-IR which restricts them to the shaded region (only 8\% (2) of the AGN sources in the grey region have a visible stellar bump). Of the outlying high redshift AGN in this plot, the majority (70\%) possess a weak visible stellar bump which pushes them into the regions of the color space occupied by SF galaxies. 
Eleven (50\%) of the silicate AGN, 2 (17\%) of the featureless AGN, and 1 (25\%) of the unclassifiable AGN lie outside the gray region. One of the advantages of this study is that we are able to
investigate how well the mid-IR power source determines the near- and far-IR power source. Clearly, our mid-IR dominated AGN do not uniformly have IRAC colors indicative of an AGN. Furthermore, 11 of the AGN contaminating the SF region do not have far-IR colors expressive of an AGN (see \S \ref{sec:new}).

The middle panel of Fig. \ref{sternlacy} is the color space defined by \citet{lacy2004} where the shaded region is used to identify AGN. The vast majority of our sources occupy
this region, regardless of their power source diagnosed from the mid-IR spectrum \citep[see also][]{donley2008,donley2012}. The SF sources again show a clean redshift separation as the IRAC channels sample the stellar bump. As neither subsample of AGN similarly exhibits a redshift separation, we do not plot them with 
different symbols according to redshift. The redshift tracks of the featureless AGN and silicate AGN composite SEDs occupy the same portion of the graph, so we only plot the silicate
AGN SED track in orange; the track lies in the upper portion of the graph.

We overplot both SF SED redshift tracks ($z\sim1$ in red and $z\sim2$ in purple) as
they show an interesting separation. At low redshifts ($z~<~2$), the $z\sim1$ SF SED track lies just outside the shaded region, while the $z\sim2$ SF SED lies inside it, but after $z=2$,
the tracks lie on top of each other. 
This difference in the color tracks between our $z\sim1$ and $z\sim2$ composite SEDs is likely due to the different intrinsic $L_{\rm IR}$ of each subsample. The $z\sim1$ sources are on average LIRGs ($L_{\rm IR}\sim4\times10^{11}{\rm L}_{\odot}$) and therefore have less dust to obscure the IRAC colors than our $z\sim2$ ULIRG SED composite; adding more dust to a SF galaxy will cause it to shift towards the top-left of this plot, which is exactly the shift we see between our $z\sim1$ and $z\sim2$ SED tracks. 
At $z\ge3$, the tracks approach the area occupied by the AGN, but as none of our SF sources possess a redshift this high, we are unable to determine
if our sources follow our tracks into the upper portion of the graph. Both of the SF SEDs accurately trace the redshift separation exhibited by our sources.

In the right panel of Fig. \ref{sternlacy}, we show on our composite SEDs the effective wavelengths of each of the IRAC bandwidths at redshifts 1, 2, and 3 (blue, green, and red, respectively). These bands straddle the stellar
bump over this redshift range. 
Both AGN SEDs have a power-law shape over these bands causing little evolution in IRAC colorspace. 

The high degree of contamination in these IRAC color-color diagnostics by high redshift SF galaxies motivated \citet{donley2012} to create more restrictive criteria than originally presented in \citet{lacy2004} 
and \citet{stern2005}. \citet{donley2012} uses a sample of AGN identified at optical and X-ray wavelengths to construct the new criteria, and now we are able to test these criteria using 
mid-IR spectroscopic AGN. We apply the new color cut presented in \citet{donley2012} to our sample in Figure \ref{donley}. The authors determined the gray shaded region was the most effective at
selecting out AGN with power-law spectra in the IRAC bandwidths, and indeed, all of our sources that meet the more restrictive criteria have flux densities such that
$S_{3.6}<S_{4.5}<S_{5.8}<S_{8.0}$. Based on our high redshift sources, we propose the simple IRAC color cuts of
\begin{eqnarray}
\log \left(S_{5.8}/S_{3.6}\right)>0.08\\
{\rm and} \, \, \log \left(S_{8.0}/S_{4.5}\right)>0.15
\end{eqnarray}
(solid lines in Fig. \ref{donley}) for selecting potential power-law AGN candidates, similar to the \citet{donley2012} criteria.

The IRAC color selection techniques, even using the more restrictive cuts presented in \citet{donley2012},
still miss a large fraction (39\% of the present sample) of mid-IR spectroscopically confirmed AGN, specifically most of the more obscured silicate AGN. Furthermore, such diagnostics cannot conclusively determine
if an AGN is significantly contributing to the bolometric luminosity of a galaxy.
Since an IRAC color diagnostic applied at high redshift ($z~\gtrsim$~1.3) is necessarily based on separating sources into AGN- or SF-dominated based on the shape of the spectrum in the
near-IR, such diagnostics might not be the most
desirable for determining the dominate power source of dust obscured galaxies in the mid-IR and far-IR regime.

\section{New Color-Color Diagnostics}
\label{sec:new}

Our large sample of high redshift SF and AGN sources, identified with deep mid-IR spectroscopy, and wealth of multiwavelength photometry allows us to define new color-color diagnostics that are well suited to uncovering galaxies harboring an AGN as revealed in the mid-IR spectrum, and galaxies with a bolometrically important AGN. We seek to combine multiple portions of the IR spectrum that are
probing the physical nature of each galaxy with different pieces of information. At $3.6\,\mu$m, SF galaxies at $z\sim1-2$ will have a stellar bump due to the underlying stellar
population, while this effect might be washed out in luminous AGN, producing a power-law shape. Similarly, at these redshifts, the $24\,\mu$m filter will straddle the PAH complexes
at 7.7 and $12.7\,\mu$m, which will be weakened or absent in a bright AGN. However, there is a caveat that at $z\sim1.5$, the silicate absorption feature present in some AGN spectra, falls into the $24\,\mu$m bandwidth which can produce colors that mimic SF galaxies. On the other hand, the bandwidth is large enough that we do not expect this to be a significant source of contamination. The far-IR should have a different shape based on the relative amounts of cold and warm dust emission present,
and SF galaxies will have relatively more cold dust than AGN, while AGN have an increased amount of warm dust (Paper I). Finally, the $8\,\mu$m filter, at $z\sim1-2$, covers a relatively
featureless portion of the spectrum, so when combined with filters that are tracing features, should act as a base to distinguish between AGN and SF systems.

 We used available photometry from {\em Herschel} and {\em Spitzer} ranging from $3.6-350\,\mu$m (observed) and explored
every possible color combination. In addition, we also look at where the composite SEDs lie to get a sense of where we should expect to find AGN and SF galaxies, on average. For color-color plots in which the AGN were well separated from the SF sources, we calculated the contamination rates of each region. We find that a color diagnostic spanning the full range of the IR spectrum does the best job of separating both AGN with pure power-law spectrum from $1-10\, \mu$m and AGN with silicate absorption from SF galaxies at high redshift.
We define two new color-color diagnostics which, based on having a low contamination and clarity of separation, are the optimal diagnostics to employ when a full suite of IR photometry exists.

\subsection{$S_{250}/S_{24}$ vs. $S_{8}/S_{3.6}$}
\label{sec:color1}
We find that combining longer wavelength photometry from {\em Herschel} with mid-IR photometry from {\em Spitzer} MIPS/IRAC provides the most reliable separation between our mid-IR classified AGN and SF galaxies since they probe the widest range of dust properties affected by AGN and SF activity.

Figure \ref{basic250_24} shows $S_{250}/S_{24}$ vs. $S_{8}/S_{3.6}$. Both colors cause a separation of sources with the result that the SF dominated
sources lie in the lower right portion of the graph (separated by the diagonal solid line). The SF sources separate weakly according to redshift (indicated by open and filled symbols) with the lower redshift sources lying slightly below the high redshift sources. The redshift tracks calculated from the $z\sim1$ SF SED and $z\sim2$ SF SED follow similar evolutionary paths, although again they exhibit a separation in the IRAC color, $S_{8}/S_{3.6}$, with the $z\sim2$ SF track redder than the $z\sim1$ SF track between $z\sim1-2$. The SED redshift tracks trace out the area occupied by the SF galaxies according to the redshift separation exhibited by the sources. 

The AGN are loosely separated by the presence
or absence of silicate absorption at $9.7\,\mu$m in this color-color space. We plot both subsamples according to redshift (filled symbols are $z\sim1$ and open symbols are $z\sim2$).
The featureless AGN do not contaminate the SF region and lie further to the left (lower $S_{250}$/$S_{24}$) than the silicate AGN, due to an excess of warm dust emission 
in these sources. We overplot the redshift track computed
from the featureless AGN SED. The track has some slight evolution along the x-axis with redshift, consistent with the individual data points.
The silicate AGN (filled and
open squares, according to redshift), on the other hand, show no evolution along the x-axis with redshift but show some spread along the y-axis. The fact that some silicate AGN sources contaminate the SF region can be attributed to two effects, namely that some individual sources possess a weak stellar bump and some sources have more relative cold dust emission in the far-IR. The silicate AGN SED redshift track exhibits no evolution in either direction and
so is plotted as a star.

Both color axes produce a separation of AGN and SF sources. Based on the location of our sources, we calculate the separation line to be
\begin{equation}
\log(S_{8}/S_{3.6}) = 0.74\times\log\left(S_{250}/S_{24}\right)-0.78 
\end{equation}
drawn as the bold solid line in Fig. \ref{basic250_24}. In the absence of all four wavelengths, we find that the majority of our AGN sources satisfy either of the following color criteria shown as the dashed lines:
\begin{eqnarray}
\log(S_{250}/S_{24}) < 1.30  \\
\log(S_{8}/S_{3.6}) > 0.32 
\end{eqnarray}  
The separation along individual axes is particularly useful if only an upper limit for $S_{250}$ is available when searching for AGN.

We quantify the contamination of the SF and AGN regions (separated by the diagonal solid line):
10\% of the sources (11 of 111 total) in the SF region are AGN and
2 of the 22 sources (9\%) in the AGN region are SF dominated. None of the contaminating AGN are power-law dominated. Some of the individual silicate AGN have a stellar bump (41\%), causing contamination along the 
$S_{8}/S_{3.6}$ axis. Futhermore, several of our silicate AGN (8, or 42\%) have significant amount of cold dust emission in the far-IR, producing $S_{250}/S_{24}$ flux density ratios greater than 1.3. 
The high  $S_{250}/S_{24}$ flux density ratio correlates with the presence of a stellar bump in the silicate AGN, with 32\% possessing both.
Finally, it is worthwhile
to note that the contaminating AGN are significantly fainter at $24\,\mu$m (median $S_{24}=220\,\mu$Jy) than the AGN lying in the upper region of the graph (median $S_{24}=1240\,\mu$Jy). Indeed, all
but one of the properly identified AGN have $S_{24} > 300\,\mu$Jy. Based on our sample of 24\,$\mu$m bright sources, our diagnostic is optimized to select AGN with $S_{24} > 300\,\mu$Jy, though if the physics is similar in lower luminosity AGN, they will have similar SEDs and colors, in which case our diagnostics can also be used to select out AGN dominated sources.

$S_{250}$, in comparison with $S_{24}$, is an indicator of the relative amount of cold dust in a galaxy.
As an AGN becomes more powerful, the relative amount of warm dust increases, so that the ratio of $S_{250}/S_{24}$ decreases. For galaxies lacking a significant amount of warm dust, the cold dust will be the dominant contributor to the $L_{\rm IR}$. Low $S_{250}/S_{24}$ ratios in our mid-IR AGN indicate that the warm dust, indicative of an AGN, is significantly contributing to the far-IR emission, and accordingly, the AGN  is an important contributor to the bolometric luminosity. Our diagnostic is therefore more powerful than the IRAC diagnostics presented in \S \ref{sec:irac} since it selects AGN significantly contributing to the total IR emission.

\begin{figure*}[ht!]
\centering
\includegraphics[scale=0.37]{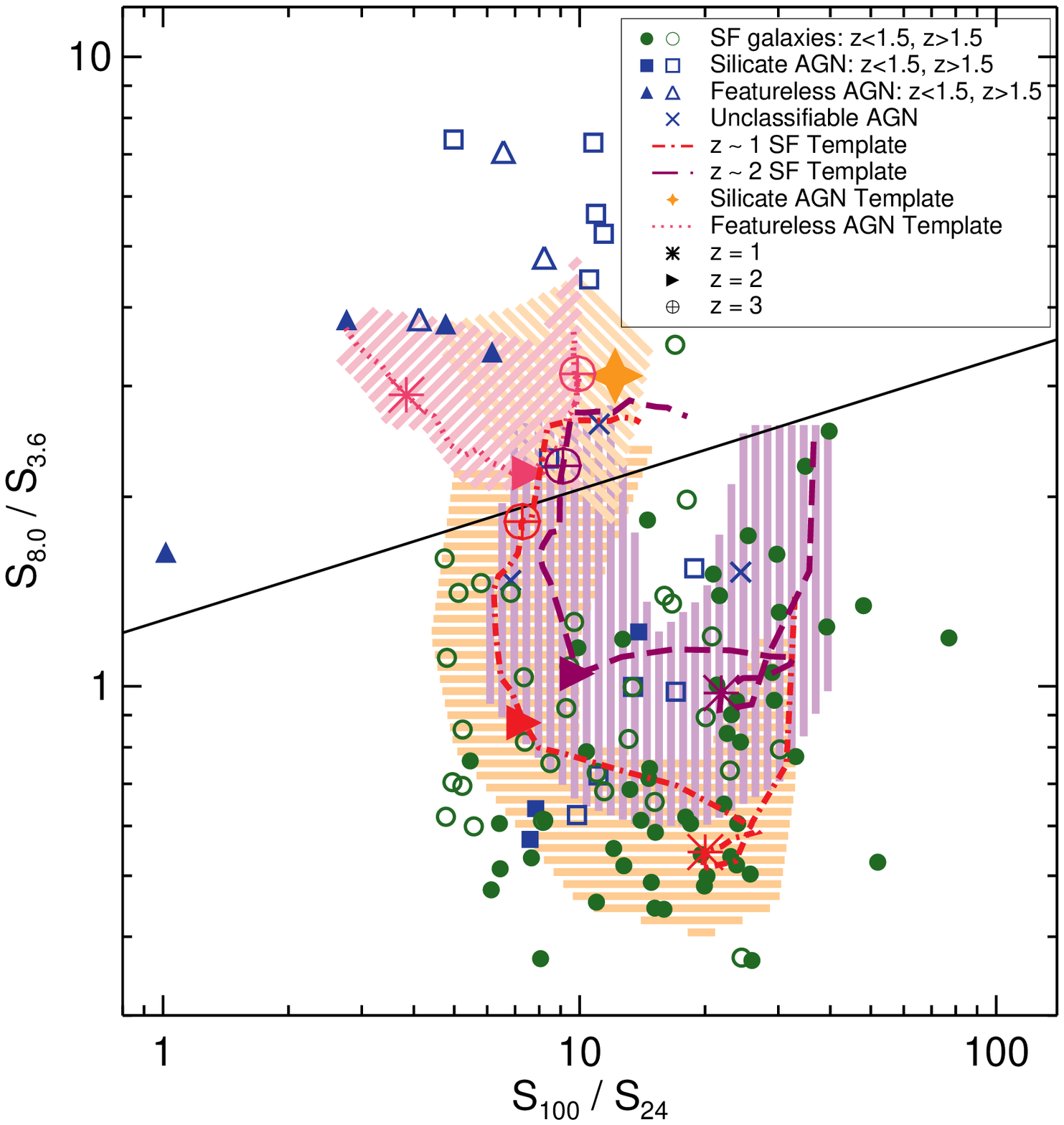}
\includegraphics[scale=0.365]{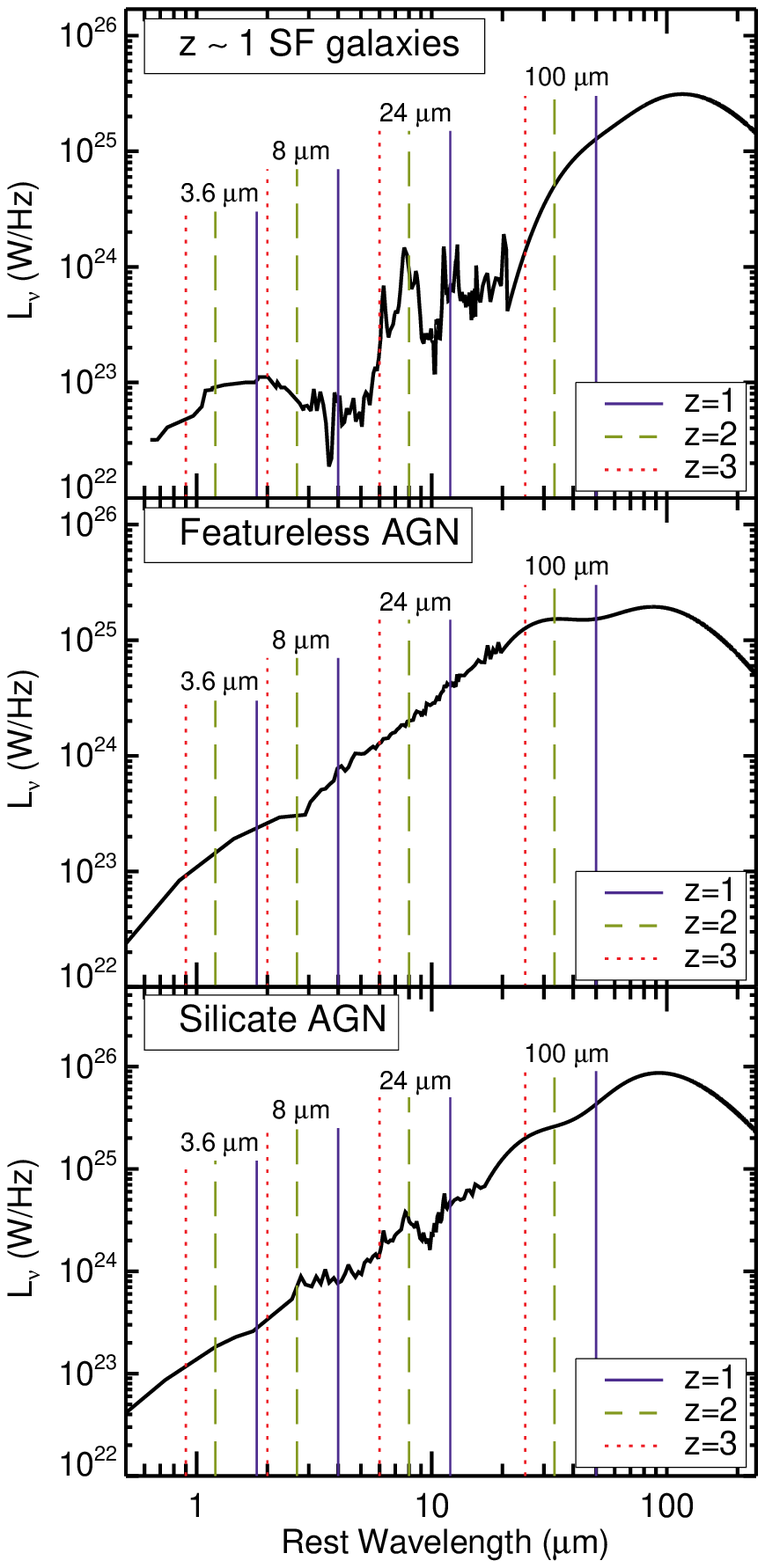}
\includegraphics[scale=0.37]{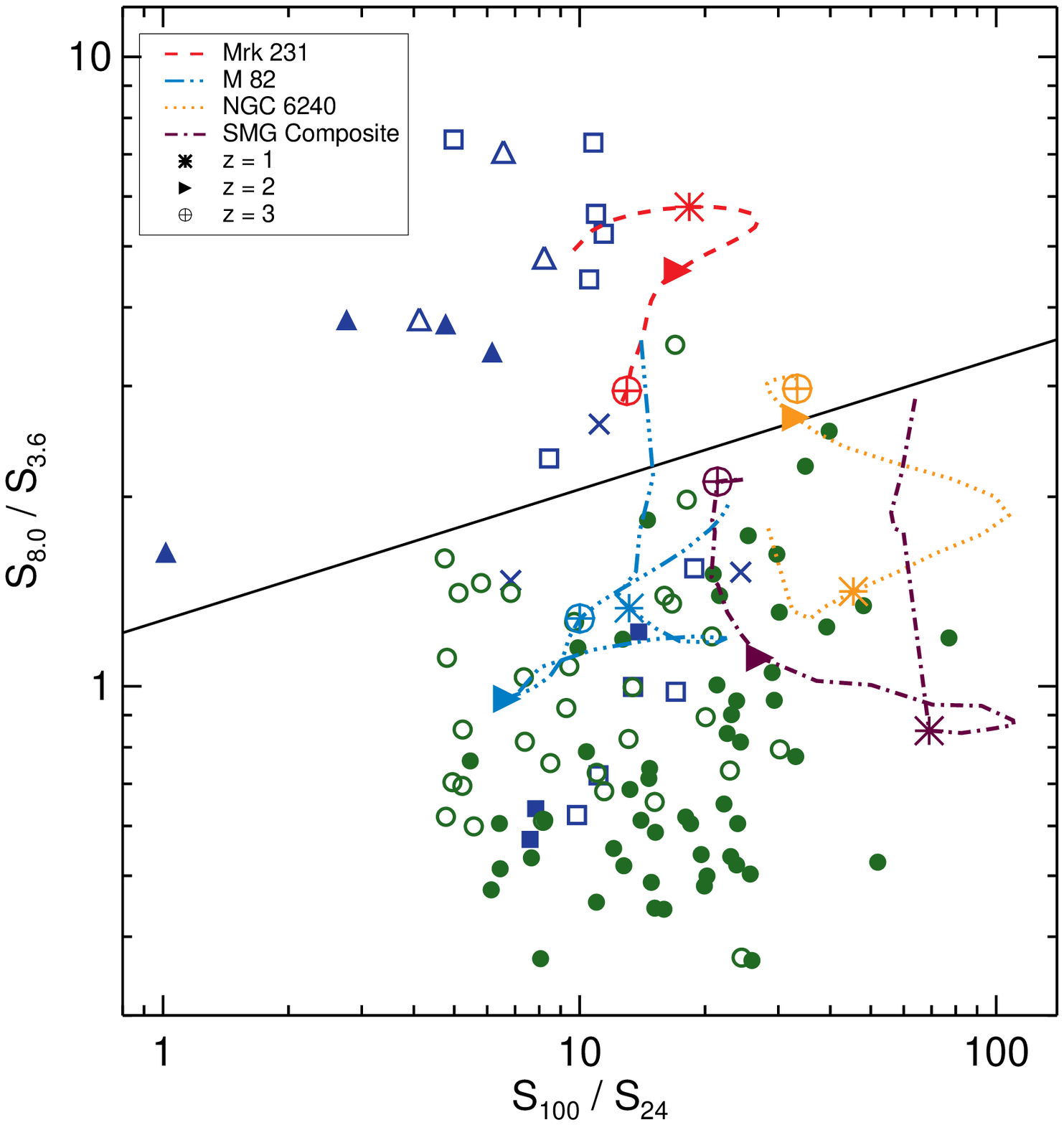}
\caption{{\em Left -} New color-color diagnostic combining far-IR {\it Herschel}/PACS photometry with mid-IR {\em Spitzer} data. The dark line divides the SF galaxies (green) from the AGN galaxies (blue). 
We plot all sources according to redshift (filled symbol: $z<1.5$; open symbol: $z>1.5$). 
We also plot the redshifts tracks of our high redshift composite SEDs from $z\sim0.5-4.0$ which is the redshift range of our sources. 
The silicate AGN composite SED has negligible redshift evolution in this color space and is plotted as the orange star.  
{\em Middle -} Illustrates where each of the redshifted photometry points lie on the composite SEDs. 
{\em Right -} Our new color-color plot with templates from local galaxies and high redshift SMGs overplotted. 
\label{basic100_24}}
\end{figure*}

Our new diagnostic is a definite improvement over the IRAC selection criteria presented in \citet{lacy2004} and \citet{stern2005}, though photometry is scarcer. In our sample of 151 galaxies, 99\% have IRAC data so both \citet{lacy2004} and \citet{stern2005} can be applied,
while only 88\% have the relevant wavelengths to satisfy our new diagnostic. We verify the more restrictive IRAC criteria of \citet{donley2012}, shown in Fig. \ref{donley}, with our IRS sample, whose mid-IR power source has been determined via spectral decomposition.
Our new diagnostic is a slight improvement over the revised IRAC criteria presented in \citet{donley2012}. In the present case, 35\% of our AGN sources are misclassified as SF galaxies, while according to the restricted IRAC criteria (equations (1) and (2)), 39\% are misclassified. In addition, with the new criteria, we correctly identify two sources as AGN that were misclassified using
only the IRAC colors. Though the improvements gained in recovering mid-IR AGN by combining colors covering the entire IR spectrum is only slight over using IRAC colors alone, an added strength of our new
criteria lies in the fact that mid-IR data from the Wide-field Infrared Survey Explorer (WISE) can be easily substituted for {\em Spitzer} data. The WISE 22\,$\mu$m and 3.4\,$\mu$m
channels correspond to {\em Spitzer} 3.6\,$\mu$m and 24\,$\mu$m bandwidths. We have used the redshift tracks of our templates to determine that the WISE 12\,$\mu$m channel is the
optimal substitute for {\em Spitzer} 8\,$\mu$m (see \S \ref{sec:wise}).

The middle panel of Fig. \ref{basic250_24} illustrates the different SED tracks shown on the left panel. $250\,\mu$m, $24\,\mu$m, $8\,\mu$m, and $3.6\,\mu$m at $z=1$, 2, 3
are indicated on the composite SEDs.
$S_{250}$ traces the far-IR peak of each template while $S_{24}$ traces the mid-IR emission -- the different ratios of mid-IR to far-IR emission (or warm to cold dust) in AGN and SF sources produces the observed
color separation in $S_{250}/S_{24}$.  At the lower wavelengths, $S_{8}/S_{3.6}$ remains fairly constant for both high redshift AGN SED templates. As some of our individual AGN sources do show signs of weak SF activity as well as AGN activity, several of our AGN possess a noticeable stellar bump which causes a spread in $S_{8}/S_{3.6}$. Of the 11 AGN sources contaminating the SF region, 60\% possess a visible stellar bump compared with only 14\% of AGN sources in the AGN region.
The $z\sim1$ SF composite SED illustrates how a stellar bump would cause a change in $S_{8}/S_{3.6}$ with redshift.

In the right panel of Fig. \ref{basic250_24}, we overplot other SED templates in our new color-color space, namely the SEDs of local galaxies and high redshift submillimeter galaxies \citep[SMGs,][]{pope2008}. These local SEDs come from combining all known data on these sources including available IRAS (12, 25, 60, 100$\,\mu$m), {\em Spitzer} (24, 70, 160$\,\mu$m), and SCUBA (850$\,\mu$m) photometry as 
well as mid-IR spectroscopy \citep[e.g.,][]{schreiber2003,armus2007}.
Mrk~231 (red) lies in the AGN region we defined, although contrary to our high redshift AGN SED templates,
it does exhibit some evolution along the y-axis. M~82 (blue) does not enter the SF region until $z=1$ which is most likely due to the fact that it's SED peaks at a lower IR wavelength than the majority
of our high redshift sources (see Paper I). Both NGC~6240 (orange) and the high redshift SMG composite (purple) lie in the SF region except at very low redshifts ($z\lesssim0.5$) and follow the same general redshift 
evolution as our sources, that is, increasing $S_{250}/S_{24}$ color with increasing redshift. 
The SMG composite reaches even higher $S_{250}/S_{24}$ colors
than our sample which is consistent with their selection at submm wavelengths. 
The consistency of these other local and high redshift templates with our new color diagnostics reinforces our confidence in applying this color selection to a wider range of IR luminous, $24\,\mu$m bright galaxies at high redshift.

\subsection{$S_{100}/S_{24}$ vs. $S_{8}/S_{3.6}$}
\label{sec:color2}


In the absence of longer wavelength SPIRE data, we find that we can substitute $S_{100}$ for $S_{250}$ and we still see a nice separation between the SF and AGN galaxies. 
In the left panel of Figure \ref{basic100_24}, we use the colors $S_{100}/S_{24}$ and $S_{8}/S_{3.6}$ to define a region that separates our high redshift SF and AGN dominated galaxies (diagonal solid line).
The line of separation is
\begin{equation}
\log(S_{8}/S_{3.6}) = 0.208 \times \log(S_{100}/S_{24}) + 0.105
\end{equation}
where mid-IR AGN lie above the line.

We plot the uncertainties on our redshift tracks as the hashed lined regions. We opt to not to plot these uncertainties in the other plots presented in this work (Figs. 
\ref{sternlacy} and \ref{basic250_24}) for the sake of clarity, and the ranges covered by the uncertainties in this plot are indicative of the spread in the previous figures.
As discussed in detail in Paper I, the uncertainties on the composites were calculated by a bootstrapping technique, which indicates how the scatter in the data points
affects the calculated median luminosity by resampling with replacement. The uncertainties on our composites are not calculated directly from the intrinsic scatter in the
data, but are the standard deviation of the calculated luminosity after resampling the data 10,000 times. Therefore, it is not surprising that uncertainties on the template tracks do not encompass the full spread of all data points, particularly the silicate AGN.
 
The SF region has only a 12\% contamination (11 of 92 total galaxies) by AGN sources. 
The SF dominated systems show a clear separation with redshift.  We overplot the redshift track
of the $z\sim1$ SF composite SED in red, and it traces out the evolution exhibited by our individual sources. 
There is only one SF source that lies
inside the AGN region (causing a contamination rate of 7\%), and it not only has a high redshift ($z~=~2.57$) but also has a 47\% AGN contribution to the mid-IR. 
The silicate AGN exhibit no redshift evolution in this color space, but the featureless AGN display a weak separation.
The AGN composite SED tracks (orange and pink) do not move much vertically with redshift since these sources have a simple power-law shape and lack a stellar bump in the rest-frame near-IR. 

The middle panel of Fig. \ref{basic100_24} shows three of our composite SEDs with lines illustrating where the relevant photometry bandwidths lie at a given redshift. The featureless AGN have a relatively flat spectrum from $20-100\,\mu$m, so the $100\,\mu$m flux does not change with redshift whereas the $24\,\mu$m flux
decreases, causing the evolution along $S_{100}/S_{24}$. This is not the case for the silicate AGN. The slope from the mid-IR to the far-IR is relatively constant, producing little
change in $S_{100}/S_{24}$ at increasing redshifts.

We have presented a simple IR color-color plot that separates our high redshift AGN and SF sources. 
In the right panel of Fig.
\ref{basic100_24}, we test our diagnostic by overplotting the redshift tracks of local templates and a high redshift SMG SED.
The local AGN Mrk~231 (red dashed line) lies well outside the SF region while the starburst M~82 (blue) and local ULIRG NGC~6240 (orange)
lie inside it for the most part. The SMG redshift track (purple) also lies in the SF region as expected since most SMGs are star formation dominated \citep[e.g.,][]{pope2008}. 

Substituting $S_{160}$ instead of $S_{100}$ also works well for separating out the AGN (it has the same contamination rates mentioned above). However, the SF sources do not have a strong
redshift separation, although, since $S_{160}$ is still probing the cold dust at lower redshifts, there is a stronger separation along the x-axis. The $S_{100}/S_{24}$ vs. $S_{8}/S_{3.6}$ colorspace can be used to select SF galaxies based on redshift as well as looking for AGN. The $z < 1.5$ SF galaxies have higher $S_{100}/S_{24}$ ratio since the $100\,\mu$m filter is tracing the cold dust. At redshift of 2, the $100\,\mu$m
filter is now tracing the warm dust, and the $24\,\mu$m flux density is boosted by the $8\,\mu$m PAH complex, producing a lower $S_{100}/S_{24}$ ratio, similar to what is seen for the AGN. The similar $S_{100}/S_{24}$ colors for both $z~2$ SF galaxies and AGN keeps this color alone from being an accurate indicator of the bolometrically important power source, unlike $S_{250}/S_{24}$.

\begin{table}
\begin{center}
\caption{Reliability of our new color-color diagnositics.\label{colortbl}}
\begin{tabular}{c|cc|cc}
\hline\hline
Color  & \multicolumn{2}{c|}{\# of galaxies}  & \multicolumn{2}{c}{\% (\#) Contamination} \\
\cline{2-3}\cline{4-5}
Diagnostic  & SF & AGN  & SF region & AGN region \\
\hline
$S_{250}/S_{24}$ v $S_8/S_{3.6}$  & 102 & 31  & 10 (11) & 9 (2) \\
$S_{100}/S_{24}$ v $S_8/S_{3.6}$  & \ 83 & 24  & 12 (11) & 7 (1)
\end{tabular}
\end{center}
\end{table}

\begin{figure}
\plotone{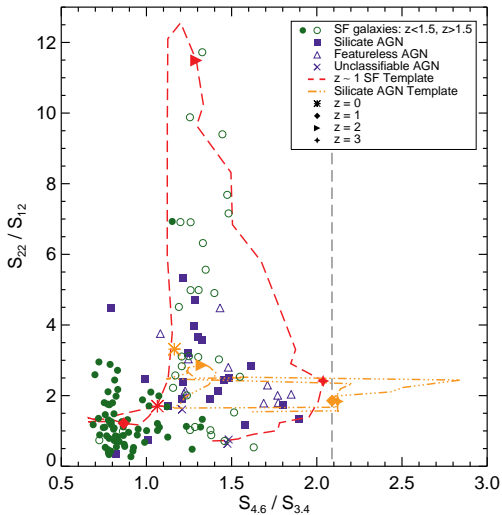}
\caption{We used the WISE transmission filters to create synthetic 12 and $24\,\mu$m photometry for our sources which we combine with {\it Spitzer} photometry of comparable 
wavelengths to create WISE colors. We overplot the redshift tracks of $z\sim1$ SF SED and silicate AGN SED. The $z\sim2$ SF SED and featureless AGN SED occupy the same
regions as the plotted redshift tracks, so for clarity we do not show them. We plot the AGN selection criterion of \citet{stern2012} as the grey dashed line. Based on the location of our 
tracks and synthetic photometry, we conclude that the primary strength of this mid-IR 
diagnostic is in selecting $z\sim2$ SF galaxies.\label{wise}}
\end{figure}

\subsection{Far-IR Color Selection}
As an AGN becomes luminous enough to dominate the mid-IR spectrum, it can heat the dust in the host galaxy causing a shift in the SED to warmer average dust temperatures
and an increased importance of warm dust to the bolometric luminosity. Based on this, we might expect that just the {\em Herschel} PACS and SPIRE colors can can be used to preferentially
select AGN sources. \citet{hatz2010} searched for a separation using $S_{350}/S_{250}$ and $S_{500}/S_{350}$ and found that in the SPIRE bands, their sample of AGN were indistinguishable
from the non-active star forming galaxies. As many of our sources are not detected or blended at $500\,\mu$m, we instead combine $S_{350}/S_{250}$ with
$S_{160}/S_{100}$ and also find that it does not separate the mid-IR classified AGN and SF sources. 
Fig. \ref{composites} illustrates that at rest frame wavelengths greater than 40$\,\mu$m, the far-IR portion of the silicate AGN SED and
both SF SEDs are all broadly consistent in shape explaining the lack of spread in colors. 

The less pronounced differences in far-IR portions of the composite SEDs is also reflected by the failure of S$_{250}$ thresholds alone to preferentially select AGN from the larger GOODS survey of galaxies
(discussed below; see Table \ref{fieldtbl1} and \S \ref{sec:field}). Our AGN 
sources, particularly the featureless AGN, are significantly
brighter than our SF sources at $24\,\mu$m and $8\,\mu$m. 
Therefore, it is not surprising that we found the largest separation between the AGN and SF sources when combining mid-IR with far-IR photometry. We caution that with only far-IR information it is 
difficult to determine the impact of the AGN on the full IR SED. The most reliable selection of AGN candidates come from combining data from $3.6\,\mu$m to $250\,\mu$m (\S \ref{sec:color1}, \S \ref{sec:color2}), which provides the most detail
about the shape of IR spectrum.

\subsection{Mid-IR Color Selection}
\label{sec:wise}
There is now an abundance of mid-IR photometry as a result of
two prominent mid-IR space telescopes: {\em Spitzer} and the Wide-field Infrared Survey Explorer (WISE). Past studies have explored using {\it Spitzer} color combinations to separate 
out mid-IR selected AGN and SF systems \citep[e.g.,][]{ivison2004, pope2008}, and emerging studies show that WISE colors can effectively separate IR luminous AGN \citep{eisenhardt2012, stern2012, yan2012}. We are motivated by these studies to investigate how well the WISE photometry can separate our high
redshift sources, particularly the SF galaxies. WISE has four transmission filters centered at 3.4, 4.6, 12, and 22\,$\mu$m, and though WISE photometry is less sensitive than {\it Spitzer} data, it has the advantage that it is an all-sky survey and can be used to search for high-redshift objects in
regions of the sky not previously well-studied.

We do not have WISE photometry for our sources, but we can use the WISE transmission
filters to create synthetic photometry using the IRS spectra and the appropriate transmission filters at 12 and $22\,\mu$m. For the 3.4 and 4.6$\,\mu$m filters, we substitute the
appropriate IRAC photometry. We have applied a small correction to the IRAC photometry (0.94 for the $3.6\,\mu$m filter and 1.04 for the $4.5\,\mu$m filter), which we have calculated
by comparing the responses of our composite SEDs to each of the WISE and IRAC transmission filters. In addition, we calculate the redshift tracks of our composite SEDs by convolving with the appropriate WISE filters, and we plot
our synthetic photometry and redshift tracks in Figure \ref{wise}.

Colors combining the first three channels, 3.4, 4.6, and 12\,$\mu$m, are capable of selecting hyperluminous
($L_{\rm IR} > 10^{13}\,{\rm L}_{\odot}$) galaxies, particularly AGN/QSOs \citep{eisenhardt2012, stern2012, yan2012}.
\citet{stern2012} found that the WISE color cut  $[3.4]_{\rm (AB)} -[ 4.6]_{\rm(AB)} > 0.8$ separates luminous AGN, which have been classified as such based on meeting the criteria
in \citet[see Fig. \ref{sternlacy}]{stern2005}. As a complement to these techniques, we would like to separate luminous SF galaxies at high redshift. We find that only combining the first three WISE channels does not effectively separate our SF sources. The strongest separation of our tracks is produced by combining
all four WISE channels as shown in Fig. \ref{wise}. 
We overplot the AGN criterion of  $[3.4]_{\rm (AB)} -[ 4.6]_{\rm(AB)} > 0.8$ as the grey dashed line. Our AGN redshift track confirms the \citet{stern2012} criterion for AGN
with $z\sim0.5-1.0$ and $z\sim3$. However, none of our sources have a power-law slope steep enough in the appropriate wavelength range to meet this criterion.

Fig. \ref{wise} illustrates that the strongest separation of our sources and redshift tracks is for SF galaxies at $z\sim2$ due to prominent PAH features lying in the $22\,\mu$m
transmission filter. The sensitivity depths at 12 and $22\,\mu$m are $\sim1$ and $\sim6\,$mJy, respectively \citep{wright2010}. Given the sensitivity limits, at a redshift of $\sim2$,
a SF galaxy would need to
be at least a ULIRG and probably even a hyper-LIRG ($L_{\rm IR} > 10^{13}\,{\rm L}_{\odot}$) to be detected by WISE. A benefit of this color selection technique is that few, if any,
extragalactic sources with luminosities less than the ULIRG threshold should occupy the same region as the $z\sim2$ ULIRGs. Therefore, our WISE color selection method is useful for selecting the brightest
SF galaxies at $z\sim2$.
In the absence of WISE photometry, {\em Spitzer} photometry can be used in similar combinations, with either $8\,\mu$m or $16\,\mu$m substituting for
$S_{12}$ \citep[e.g.,][]{ivison2004,pope2008}.

\begin{figure*}[ht!]
\centering
\includegraphics[scale=0.58]{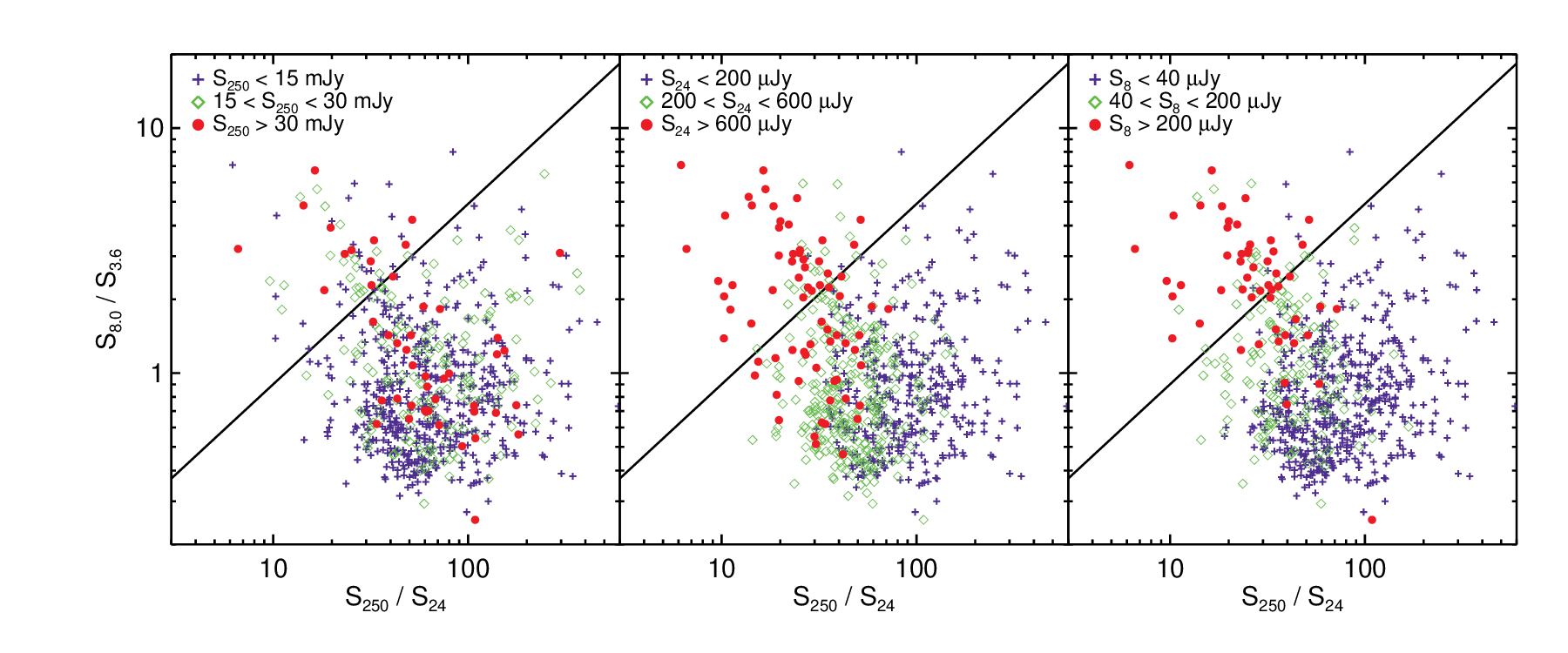}
\includegraphics[scale=.58]{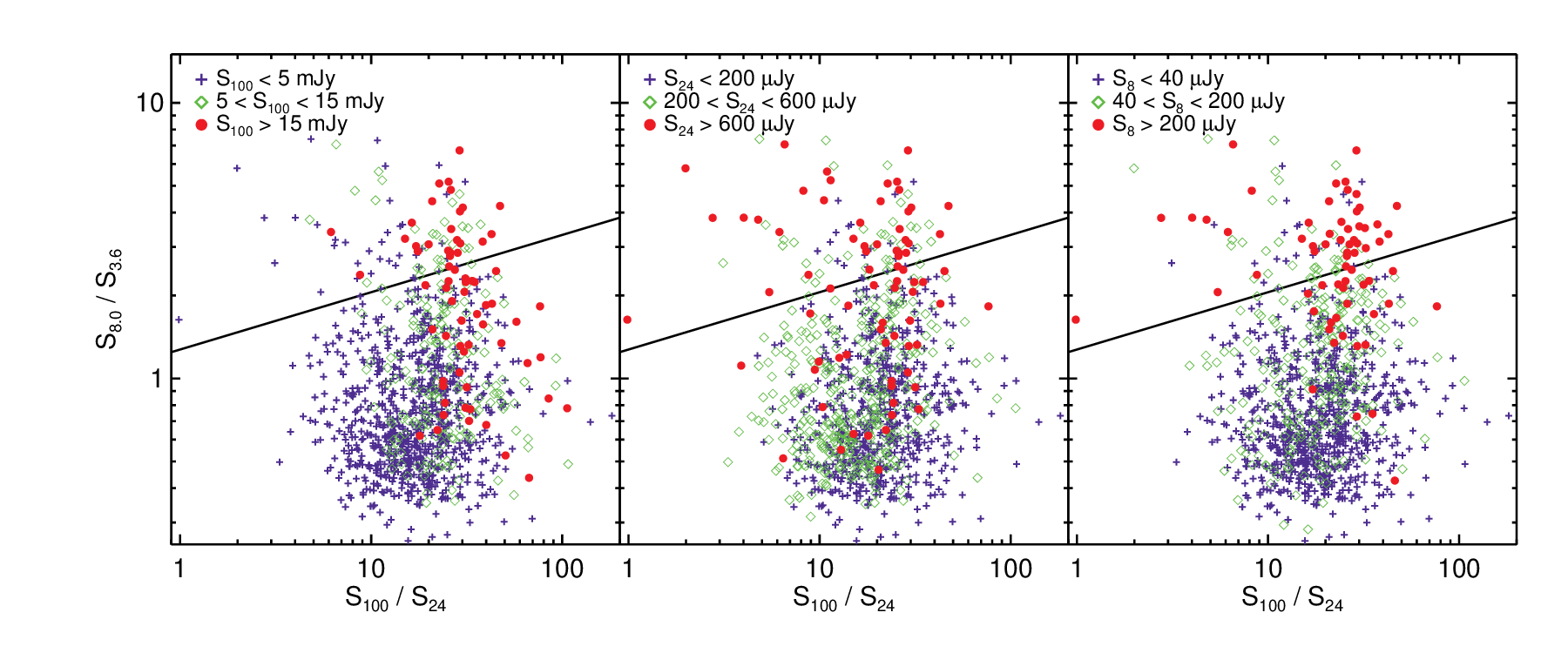}
\caption{
Our new color diagnostic for separating AGN and SF galaxies presented in Figs. \ref{basic250_24} and \ref{basic100_24} (top and bottom panels, respectively) applied to our full sample of MIPS $24\,\mu$m selected galaxies in the GOODS fields which have
corresponding {\it Herschel} photometry taken in the GOODS-{\it Herschel} survey. 
In the left panels, we color code galaxies according to their $250\,\mu$m flux density (or 100$\,\mu$m flux density); in the middle panels,
according to $24\,\mu$m flux density; and in the right panels, according to $8\,\mu$m flux density. 
The region below the solid lines are the same as we defined in Figs. \ref{basic250_24} and \ref{basic100_24};
a source lying above the line we label as AGN-dominated in the mid-IR (based on our extensive mid-IR spectroscopic sample). 
Our new diagnostics indicate that roughly $\sim$8\% of the GOODS-{\it Herschel} galaxies have an AGN dominating the mid-IR luminosity. 
Breaking things down as a function of flux density, we find that sources with the brightest $S_{24}$ and $S_8$ are predominantly AGN, whereas $S_{250}$ and $S_{100}$ do not preferentially
select AGN at brighter flux densities (see Tables \ref{fieldtbl1} and \ref{fieldtbl2}). 
\label{field}}
\end{figure*}

\subsection{X-ray emission in AGN}
One of the motivating reasons for IR diagnostics of a galaxy's power source is to search for moderately to heavily obscured AGN missed by X-ray surveys. We use {\em Chandra} surveys of the GOODS fields
\citep{alexander2003, luo2008} to determine if our diagnostics select galaxies that are obscured and lack an X-ray detection.
For $S_{250}/S_{24}$ vs. $S_{8}/S_{3.6}$, the X-ray detection fraction is 75\% for AGN lying above the solid line and drops to 36\% for AGN lying in the region dominated by SF galaxies.
For $S_{100}/S_{24}$ vs. $S_{8}/S_{3.6}$, the X-ray detection fraction for AGN above the solid line is 79\%, comparable to the detection fraction
for the $S_{250}/S_{24}$ vs. $S_{8}/S_{3.6}$ diagnostic. However, in the SF region, the X-ray detection fraction of AGN sources is 60\%. Approximately 40\% of the X-ray undetected AGN
lie above the solid lines in both diagnostics, making our color selection criteria useful
for selecting galaxies dominated by an AGN in the IR regardless of the level of obscuration. It is interesting to note that $\sim60\%$ of our mid-IR dominated AGN would not be classified as such based on
either X-ray data or the $S_{250}/S_{24}$ and $S_{8}/S_{3.6}$ colors. The AGN embedded in these galaxies are not having a strong enough effect on the far-IR and near-IR portions of the spectrum to distinguish them
from SF galaxies on the basis of colors, although the AGN are having an effect on the mid-IR emission. This could be a product of viewing angle, or possibly even an evolutionary sequence
where as the AGN grows more luminous, it will be reflected in its IR colors (see Paper I for a full discussion).

\section{Application of new color-color diagnostic to all GOODS-{\it Herschel} galaxies}
\label{sec:field}
\begin{table*}[ht!]
\begin{center}
\caption{Percentage of AGN and SF field galaxies according to S$_{250}$, S$_{24}$, and S${_8}$ strength. \label{fieldtbl1}}
\begin{tabular}{ccc|ccc|ccc}
\hline\hline
\multicolumn{3}{c}{S$_{250 \mu m}$} & \multicolumn{3}{c}{S$_{24 \mu m}$} & \multicolumn{3}{c}{S$_{8 \mu m}$} \\
\hline
Range (mJy) & \% AGN (\#) & \% SF (\#) & Range ($\mu$Jy) & \% AGN (\#) & \% SF (\#) &Range ($\mu$Jy) & \% AGN (\#) & \% SF (\# )\\
\hline
$<$ 15  & \ 5 (25) & 95 (480) & $<$ 200 & \ 1 (3) & 99 (352) & $<$ 40  & \ 1 \ (6) & 99 (497) \\
15-30   & 12  (21) & 88 \ (150) & 200-600 & \ 6 (19) & 94  (278)  & 40-200  & 10 (17) & 90  (151) \\
$>$ 30 & 26  (12) & 74  \ (35)  & $>$ 600 & 51 (36) & 49 \ (35)   & $>$ 200 & 67 (35) & 33  \ (17)
\end{tabular}
\end{center}
\tablecomments{Data corresponds to the top panels of Fig. \ref{field} which shows GOODS-N and GOODS-S field galaxies plotted on our color-color diagnostic defined in Fig. \ref{basic250_24}. AGN 
are those lying above the solid line indicated on the plot. In this
table, we calculate the percentage (number) of AGN and SF sources in a given flux density range for S$_{250}$, S$_{24}$, and S$_{8}$.}
\end{table*}

We now apply our new diagnostics defined above to broadly separate SF and AGN dominated galaxies in the whole GOODS-{\em Herschel} survey \citep{elbaz2011}.
We plot all galaxies in GOODS-N and GOODS-S detected in all four bands on our two new color-color plots in Figure \ref{field}. We use the color cuts derived in Sections \ref{sec:color1} and \ref{sec:color2} (equations (3) and (6)) to separate sources dominated by AGN and SF activity in the mid-IR.
Using the $S_{250}/S_{24}$ vs. $S_8/S_{3.6}$ plot (top panels), we have a total of 665 galaxies with detections in all four bandwidths in GOODS-N and GOODS-S, of which 58 (9\%) lie in the AGN region.
For $S_{100}/S_{24}$ vs. $S_8/S_{3.6}$ (bottom panels), we have 988 sources with detections in all four badwidths, of which 94 (10\%) lie in the AGN region. Both diagnostic plots lead to a similar fraction of GOODS-{\it Herschel} sources being AGN dominated in the mid-IR. 

It has been found that bright $24\,\mu$m flux density correlates with mid-IR AGN indicators at high redshift and in local ULIRGs \citep[respectively]{desai2007,dey2008,donley2008}.
Given this trend, we look to quantify the fraction of galaxies that are AGN dominated in the mid-IR as a function of flux density using our new color-color plots. 
In Fig. \ref{field}, we plot all GOODS-{\it Herschel} galaxies with different colored symbols depending on their flux density (see legend for each panel). Our diagnostic was determined using
$24\,\mu$Jy bright sources, and applied to our own sample, the diagnostic primarily recovered AGN with $S_{24}>300\,\mu$Jy. Furthermore, the contribution of the AGN to the mid-IR has been shown to increase 
with
increasing 24\,$\mu$m flux density, particularly at higher redshift \citep[$z>0.6$][]{brand2006}.
Therefore, it is perhaps not surprising that only one fainter source ($S_{24}<200\,\mu$Jy) is found in the AGN region in Fig. 7.
It is clear that a larger fraction of sources are in the AGN region of the color-color plot for the brightest 24$\,\mu$m and 8$\,\mu$m flux densities, whereas the $250\,\mu$m and $100\,\mu$m flux densities do not make as much of a difference to the AGN fraction. We quantify the fraction of sources in each flux bin that are classified as AGN and SF dominated based on these color-color plots in Tables \ref{fieldtbl1} and \ref{fieldtbl2}. Our results show that the presence of an AGN is only visible as an excess in the mid-IR, while the far-IR photometry alone is largely insensitive to the physics of the nuclear power source. 
We therefore conclude that the flux limit of a mid-IR survey has a big effect on the fraction of sources which have a significant AGN whereas far-IR/submm fluxes correlate less strongly with AGN fraction.

\begin{table*}
\begin{center}
\caption{Percentage of AGN and SF field galaxies according to S$_{100}$, S$_{24}$, and S${_8}$ strength. \label{fieldtbl2}}
\begin{tabular}{ccc|ccc|ccc}
\hline\hline
\multicolumn{3}{c}{S$_{100 \mu m}$} & \multicolumn{3}{c}{S$_{24 \mu m}$} & \multicolumn{3}{c}{S$_{8 \mu m}$} \\
\hline
Range (mJy) & \% AGN (\#) & \% SF (\#) & Range ($\mu$Jy) & \% AGN (\#) & \% SF (\#) &Range ($\mu$Jy) & \% AGN (\#) & \% SF (\# )\\
\hline
$<$ 5  & \ 5 (40)  & 95 (779) & $<$ 200  & \ 3 (22)  & 97 (613) & $<$ 40  & \ 3 (21)  & 97 (741) \\
5-15    & 15 (29)  & 85 (165) & 200-600  & \ 9 (35)  & 91 (337) & 40-200  & 13 (33)  & 87 (222) \\
$>$ 15 & 36 (25) & 64 \ (44)  & $>$ 600  & 49 (37) & 51 \ (38)  & $>$ 200 & 62 (40) & 38 \ (25)
\end{tabular}
\end{center}
\tablecomments{Data corresponds to the bottom panels of Fig. \ref{field} which shows GOODS-N and GOODS-S field galaxies plotted on our color diagnostic presented in Fig. \ref{basic100_24}.
AGN sources (94 total) are those lying above the solid line indicated on the plot. In this
table, we calculate the percentage (number) of AGN and SF sources in a given flux density range for S$_{100}$, S$_{24}$, and S$_{8}$.}
\end{table*}

\begin{figure}
\plotone{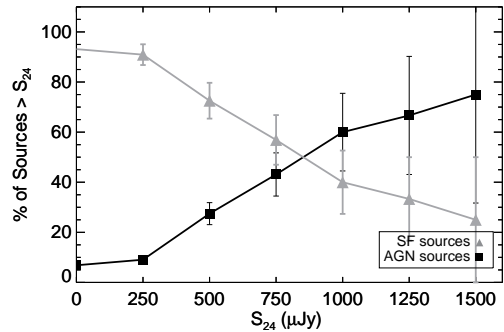}
\caption{We calculate the percentage of AGN and SF sources (with Poisson error bars) greater than a given 24\,$\mu$m flux density.
Sources are taken from the GOODS fields and defined as an SF or AGN galaxy based on their position
in $S_{250}/S_{24}$ vs. $S_8/S_{3.6}$ color space (see top panels of Fig. \ref{field}). The brightest flux densities are dominated by AGN, illustrating that applying a flux density cut in $S_{24}$ is useful for selecting
a population with a large percentage of AGN.  \label{cumfrac}}
\end{figure}

Using the data in Table \ref{fieldtbl1}, we plot the percentage of galaxies which are classified as AGN and SF greater than a given 24\,$\mu$m flux density in Figure \ref{cumfrac}.
The fraction of AGN sources shows a steady increase with 24$\,\mu$m flux density.
Above $S_{24}\sim750\,\mu$Jy, the majority of galaxies selected at these wavelengths will be AGN dominated. In Table \ref{basictbl}, we compare the intrinsic luminosities of our composite
SEDs. Despite our AGN SEDs having an $L_{\rm IR}$ below that of
the $z\sim2$ SF SED, the AGN subsamples have a higher median $S_{24}$ flux density, particularly the featureless AGN. This effect is due to the
differing ratios of far-IR to mid-IR luminosity. With the possible exception of $z\sim2$, where the brightest PAH complex falls in the $24\,\mu$m filter, at all redshifts AGN should comprise the majority of galaxies above $S_{24}>750\,\mu$m despite possibly being less intrinsically luminous in the IR than SF galaxies (see Paper I for a direct comparison of
our SEDs).

\section{Conclusions}
We have combined deep photometry from $3.6-500\,\mu$m with {\em Spitzer} mid-IR spectroscopy for 151 high redshift ($z>0.5$) (U)LIRGs in order to explore the relative positions
of high redshift mid-IR AGN and SF galaxies in IR colorspace. We start by applying color-color diagnostics based on IRAC colors alone \citep{lacy2004, stern2005} to our high redshift sample and find that there is
significant contamination of SF sources in the AGN regions; however, our mid-IR spectroscopic sample confirms the new IRAC color diagnostics from \cite{donley2012}. Adding $24\,\mu$m photometry does not effectively
separate all of the AGN, but it does produce a separation of the SF galaxies according to redshift. The same separation of SF galaxies is produced by plotting our redshifted SED tracks in color-color
graph using the WISE transmission filters.

Our high redshift AGN and SF sources exhibit different spectral features and SED shape in the near-, mid-, and far-IR/submm wavelengths. We explore combining photometry from all three IR ranges for
an optimal selection technique. In addition to photometry, we explore where AGN or SF sources should lie on average in different color combinations by using redshift tracks of our composite SEDs.
We present two new color-color diagnostics combining {\it Spitzer} mid-IR and {\it Herschel} far-IR photometry that can be used to estimate whether a galaxy harbors an AGN
in the absence of IRS spectroscopy. The optimal color diagnostics for bright $S_{24}$ sources spanning the full IR spectrum are $S_{250}/S_{24}$ vs. $S_{8}/S_{3.6}$ and $S_{100}/S_{24}$ vs. $S_{8}/S_{3.6}$. These diagnostics have a low contamination rate
of $\sim$10\% in each of the SF and AGN occupied regions, and the contaminating AGN chiefly have $S_{24} < 300\,\mu$Jy.  Moreover, the $S_{250}/S_{24}$ color estimates if an AGN is significantly contributing 
to the bolometric output of a galaxy.
The mid-IR photometry from WISE can also be used in conjunction with {\em Herschel} photometry to separate out AGN. When only limited data is available,
either of the colors $S_{250}/S_{24}$ or $S_{8}/S_{3.6}$ can be used alone to select AGN.

We apply our new color-color diagnostics to the entire GOODS-{\it Herschel} survey to determine the fraction of sources which are dominated by AGN and SF activity. We find roughly 10\% of GOODS-{\it 
Herschel} galaxies have a significant AGN. We also confirm that a higher fraction ($\gtrsim50\%$) of the brightest sources, $S_{8}>200\,\mu$Jy or $S_{24}>600\,\mu$Jy, have colors indicative of a bolometrically significant AGN. 
The $100\,\mu$m and $250\,\mu$m fluxes show a much weaker dependence on AGN fraction, though when combined with $S_{24}$, $S_{250}$ effectively separates out the AGN, indicating that the amount of cold dust emission relative to mid-IR emission is a good indicator of the IR power source. 
We conclude then that far-IR fluxes or colors alone cannot be used to determine the nature of the IR power source, and the most effective method is to combine mid-IR and far-IR data.\\

\acknowledgments
We thank the anonymous referee for the careful reading of this paper and the insightful comments provided.
This work is based in part on observations made with {\it Herschel Space Observatory}, a European Space Agency Cornerstone Mission with significant participation by NASA, and the {\it Spitzer Space Telescope}, which is operated by the Jet Propulsion Laboratory, California Institute of Technology under a contract with NASA. Support for this work was provided by NASA through an award issued by JPL/Caltech.

\end{document}